\documentclass[reprint,amsmath,amssymb,aps,prb,showpacs,floatfix,superscriptaddress]{revtex4-1}
\usepackage{graphicx}
\usepackage{bm}
\usepackage{hyperref}
\usepackage{verbatim}
\usepackage{color}
\renewcommand{\figurename}{Fig.}

\newcommand{\sectionname}{Section}

\newcommand{\e}{e}
\newcommand{\ii}{i}
\newcommand{\mc}{\mathcal}
\newcommand{\mr}{\mathrm}
\newcommand{\bb}{\mathbf}
\newcommand{\ve}{\varepsilon}

\newcommand{\pd}{\partial}
\newcommand{\dif}{\mathrm{d}}

\newcommand\abs[1]{\lvert#1\rvert}

\newcommand\avgs[1]{\langle#1\rangle}

\newcommand\bra[1]{\langle#1\rvert}
\newcommand\ket[1]{\lvert#1\rangle}

\newcommand{\Real}{\operatorname{Re}}
\newcommand{\Imag}{\operatorname{Im}}
\newcommand{\Tr}{\operatorname{Tr}}

\newcommand{\ad}{\hat{a}^{\dag}}
\newcommand{\aan}{\hat{a}^{\phantom{\dag}}}

\newcommand{\ban}{\hat{b}^{\phantom{\dag}}}
\newcommand{\Bd}{\hat{B}^{\dag}}
\newcommand{\Ban}{\hat{B}^{\phantom{\dag}}}
\newcommand{\cd}{\hat{c}^{\dag}}
\newcommand{\can}{\hat{c}^{\phantom{\dag}}}
\newcommand{\dd}{\hat{d}^{\dag}}
\newcommand{\dan}{\hat{d}^{\phantom{\dag}}}
\newcommand{\ed}{\hat{e}^{\dag}}
\newcommand{\ean}{\hat{e}^{\phantom{\dag}}}

\newcommand{\leadqn}{\ell}
\newcommand{\vg}{V_{g}} %\ve_{d}
\newcommand{\Fm}{F_{-}}
\newcommand{\Fp}{F_{+}}
\newcommand{\Fpm}{F_{\pm}}
\newcommand{\bFm}{\bar{F}_{-}}
\newcommand{\bFp}{\bar{F}_{+}}
\newcommand{\bFpm}{\bar{F}_{\pm}}
\newcommand{\dS}{S_{\delta}}
\newcommand{\bdS}{\bar{S}_{\delta}}

\newcommand{\imai}{\mathrm{i}}
\newcommand{\rhor}{\hat{\rho}}
\newcommand{\rhorme}{\rho} % \rho^S
\newcommand{\hatt}[1]{\hat{#1}}      %{#1}
\newcommand{\Lj}{\hat{L}_{j}^{\phantom{\dag}}} % \hat{L}^{j}
\newcommand{\Ljd}{\hat{L}_{j}^{\dag}}          % \hat{L}^{j\dag}
\newcommand{\Ljme}{L^{j}}                %
\newcommand{\Ljmed}{L^{j*}}              %
\newcommand{\tLjme}{\tilde{L}^{j}}       %
\newcommand{\tLjmed}{\tilde{L}^{j*}}     %
\newcommand{\tLjpme}{\tilde{L}^{j'}}     %
\newcommand{\Gamj}{\Gamma_{j}}           %
\newcommand{\rmhbar}[1]{}                % #1

\begin{document}

%-----------------------------------------------------------%
\title{A phenomenological position and energy resolving Lindblad approach to quantum kinetics}

\author{Gediminas Kir{\v{s}}anskas}
\affiliation{Mathematical Physics and NanoLund, University of Lund, Box 118, 22100 Lund, Sweden}

\author{Martin Francki{\'{e}}}
\affiliation{Mathematical Physics and NanoLund, University of Lund, Box 118, 22100 Lund, Sweden}
\affiliation{Now at: Institute for Quantum Electronics, ETH Z{\"{u}}rich, Auguste-Piccard-Hof 1, 8093 Z{\"{u}}rich, Switzerland}

\author{Andreas Wacker}
\affiliation{Mathematical Physics and NanoLund, University of Lund, Box 118, 22100 Lund, Sweden}

\date{Jan 4, 2018: Accepted by Physical Review B}
%-----------------------------------------------------------%

%-----------------------------------------------------------%
\begin{abstract}
A general theoretical approach to study the quantum kinetics in a system
coupled to a bath is proposed. Starting with the microscopic interaction, a
Lindblad master equation is established, which goes beyond the common
secular approximation. This allows for the treatment of systems, where
coherences are generated by the bath couplings while avoiding the
negative occupations occurring in the Bloch-Wangsness-Redfield kinetic
equations. The versatility and accuracy of the approach is
verified by its application to three entirely different physical systems:
(i) electric transport through a double-dot system coupled to electronic
reservoirs, (ii) exciton kinetics in coupled chromophores in the presence
of a heat bath, and (iii) the simulation of quantum cascade lasers, where the
coherent electron transport is established
by scattering with phonons and impurities.
\end{abstract}
%-----------------------------------------------------------%

\maketitle

\section{Introduction}

The dynamical behavior of quantum systems coupled to a bath is a central
question for a wide range of physical problems. The classical example is the
evolution of a spin in a time-dependent magnetic field in the presence of
thermal excitations of the hosting
material\cite{BlochPR1946}. Other examples, just to mention a few, are:
Transport of electrons through quantum dot systems, where the bath is
constituted by  connecting electron reservoirs at given temperature and
electrochemical
potential\cite{KastnerRMP1992,KouwenhovenBook1997,ReimannRMP2002,HansonRMP2007};
Kinetics of excitons in molecular aggregates with their coupling to the
vibrations\cite{AbramaviciusChemRev2009,IshizakiARCondMat2012};
Electron transport in extended semiconductor heterostructures, such as
superlattices \cite{EsakiIBM1970,WackerPhysRep2002,IottiPRB2005} or quantum
cascade lasers\cite{FaistScience1994,JirauschekApplPhysRev2014}, where the
energy relaxation due to phonon scattering is crucial.

In general, the state of the quantum system can be described by the reduced
density operator $\rhor$ of the system (which is the full density operator
after tracing out the degrees of freedom from the baths). Thus, the common
problem is to determine $\rhor$ on the basis of the system Hamiltonian
$\hatt{H}_S$ in combination with the bath properties and the specific
microscopic coupling mechanism.

In order to evaluate $\rhor$, the coupling to the baths can be treated
perturbatively and a large variety of different approaches has been suggested.
For more recent examples see Refs.~[\onlinecite{TimmPRB2008,SubotnikJCP2009,SchallerPRA2009,TajEurPhysJB2009,KollerPRB2010,EspositoJPhysChemC2010,ChenJPhysChemC2014}] and references cited therein. Starting with the unmanageable von Neumann equation of the density operator for the full system, a common strategy is to obtain a similar first-order differential equation for $\rhor$, which is local in time. In the basis of the eigenstates $\ket{a}$  for the system Hamiltonian $\hatt{H}_S$ with energies $E_a$ this equation in general reads
\begin{equation}\label{EqTensor}
\begin{aligned}
\rmhbar{\hbar}\frac{\pd}{\pd t} \rhorme_{ab}=&\imai(E_b-E_a)\rhorme_{ab}\\
&+\imai\bra{a}[\rhor,\hatt{H}_\mathrm{ext}(t)]\ket{b}
-\sum_{cd}K_{abcd}\rhorme_{cd},
\end{aligned}
\end{equation}
where $\hatt{H}_\mathrm{ext}(t)$ describes possible external excitations of
the system by time-dependent fields. We note that our units are
$\hbar=1$, $k_{B}=1$, $\abs{e}=1$ except in Sections~\ref{SecPalmieri},
\ref{SecSL}, and \appendixname~\ref{ImpScatt}. Standard perturbation theory in
the system-bath couplings provides the Wangsness-Bloch-Redfield (WBR) equations
\cite{WangsnessPR1953,RedfieldIBM1957}, and $K_{abcd}$ becomes the Redfield
tensor $K^\mathrm{Red}_{abcd}$ . However, the WBR equations do not guarantee
the positivity of probabilities, which is clearly an unphysical feature albeit
other quantities such as total currents (see Refs.~[\onlinecite{WeberPRB2009,PanPRB2017}])
are often well recovered. In fact, only a special class of first-order
differential equations specified by  Lindblad \cite{LindbladCMP1976}
and Gorini \textit{et al.} \cite{GoriniJMP1976} guarantees
the  positivity of $\rhor$. The most general differential equation for
the reduced density operator, which is local in time and which conserves positivity, is given by (see, e.~g., chapter 3.2.2
of Ref.~[\onlinecite{BreuerBook2006}]):
\begin{equation}\label{EqLindblad}
\begin{aligned}
\rmhbar{\hbar}\frac{\pd }{\pd t}\rhor=&\imai[\rhor,\hatt{H}_{\rm eff}]\\
&+\sum_{j}\Gamj\Big(\Lj\rhor\Ljd
-\frac{1}{2}\rhor\Ljd\Lj
-\frac{1}{2}\Ljd\Lj\rhor\Big).
\end{aligned}
\end{equation}
Here $\hatt{H}_{\rm eff}(t)$ contains the Hamiltonian
$\hatt{H}_{S}+\hatt{H}_\mathrm{ext}(t)$ as well as possible
renormalization terms from the couplings to the baths. The
dimensionless jump operators $\Lj$ can be chosen without further
restrictions within the Hilbert space of the system and $\Gamj$ is a
real number with dimension of energy.\footnote{There are different
  ways to write Eq.~\eqref{EqLindblad}. For example, $\Gamj$ can be
  incorporated into $\hat{L}_j$.} For the basis of eigenstates, this
provides a corresponding tensor $K_{abcd}$ in Eq.~\eqref{EqTensor},
which has special properties as discussed in
Ref.~[\onlinecite{PalmieriJCP2009}].

Removing all terms from the Redfield tensor $K_{abcd}^\mathrm{Red}$
where $E_b-E_a\neq E_d-E_c$, which is called secular approximation
(sometimes also rotating wave approximation), renders a Lindblad type
tensor $K_{abcd}^\mathrm{Sec}$ together with renormalization terms in
$\hatt{H}_{\rm eff}(t)$  \cite{BreuerBook2006}. However, in this case
the coherences $\rhorme_{ab}$ for non-degenerate levels just decay (if they are not driven externally), while the
populations $P_a=\rhorme_{aa}$ of the states are solely determined by
a Pauli master equation:
\begin{equation}
\frac{\pd}{\pd t} P_{a}=\sum_c \left(R_{c\to a}P_{c}-R_{a\to c}P_a\right),
\end{equation}
with the transition rates $R_{c\to a}$. This excludes the description of a
rich field of physics where coherences are actually generated by
the bath couplings. This is relevant for, e.g., exciton kinetics
\cite{PalmieriJCP2009}, resonant tunneling in heterostructures
\cite{WackerPRL1998,CallebautJAP2005}, and carrier
capture \cite{RosatiPRB2017}. Thus, establishing a Lindblad
master equation, where coherences are fully taken into account beyond the
secular approximation, is a matter of high interest and several proposals have
been made recently\cite{TajEurPhysJB2009,PalmieriJCP2009,RosatiPRB2014b,HodJPhysChemA2016}.

In this paper we suggest a scheme based on a  phenomenological approach, where we
require that the jump operators carry information on both
the spatial and energetic properties of the jump processes.
This Position and Energy Resolving Lindblad (PERLind) approach is straightforward
to implement and we demonstrate its versatility to a wide range of systems
covering basic transport physics, chemistry, and device technology.

The paper is organized as follows: In \sectionname~\ref{SecModel} we specify
our PERLind approach, which is based on a heuristic
argument. This section is the core of our paper, while the
  subsequent sections demonstrate three applications of the
  PERLind approach in different fields of physics, chemistry, and technology. Depending on the interest of the reader, they can be read independently of each other and highlight different technical aspects of the approach. In
\sectionname~\ref{SecDoubleDot} we consider tunneling through a quantum-dot
system, where we compare the PERLind approach
with exact results and other common
approximations such as the Pauli master equation and Redfield kinetics.
Moreover, we address the approximate fulfillment of the Onsager
relation here.  \sectionname~\ref{SecPalmieri} discusses
energy transfer in chromophores in direct comparison with a different
approach \cite{PalmieriJCP2009} addressing the same problem.
The application of the PERLind approach
to quantitative simulations of quantum cascade lasers is addressed in
\sectionname~\ref{SecSL}, where it actually provides the same
type of equations as suggested in Ref.~[\onlinecite{GordonPRB2009}].
Several technical details  including the relaxation to thermal equilibrium are
  provided in the appendices.

%%%%%%%%%%%%%%%%%%%%%%%%%%%%%%%%%%%%%%%%%%%%%%%%%%%%%%%%%%%%%
\section{Defining the position
    and energy resolving Lindblad approach}\label{SecModel}

\begin{figure}
\begin{center}
\includegraphics[width=\columnwidth]{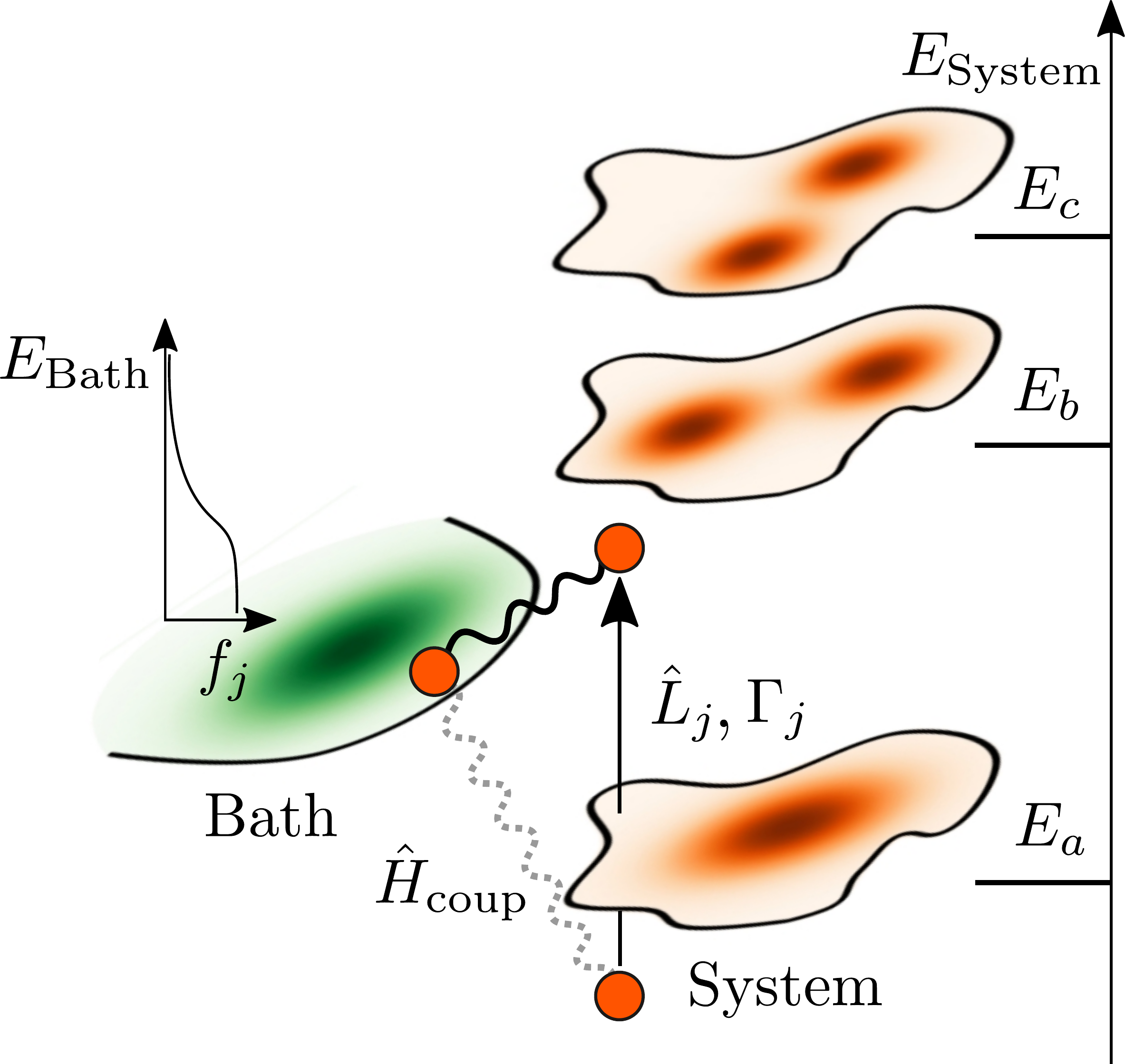}
\caption{(Color online) Scheme of the general problem addressed: The coupling
  to the bath (left) affects the quantum kinetics in a system (right), where
  three different energy eigenstates with energies $E_a$, $E_b$, and $E_{c}$ are
  depicted. Due to the spatial location of the bath, the transitions by the
  jump operator $\hat{L}_j$ occur in
  a particular region of the system as visualized by the vertical arrow. On
  the other hand, the transition strength depends on the  spectral properties
  of the bath coupling $f_j(E)$, which requires energy information of the
  system states. }
\label{fig1}
\end{center}
\end{figure}

The background for the approach is a general physical problem in the
description of interactions with the bath as sketched in Fig.~\ref{fig1}.
In many cases this interaction
requires both information on spatial and energetic properties of the
system. For example, in quantum dot systems electrons tunnel from a lead into
the region of the dot, which is adjacent to the lead. At the same time, the
lead only offers electrons with energies up to its electrochemical
potential. This energy information is contained in the eigenstates
$\phi_a({\bf r})$ of the dot, which are, however, often
extended. These two demands imply an inherent conflict:
If the tunneling process is modeled by a jump operator creating
$\phi_a({\bf r})$, the new electron would be observable at quite a distance
immediately, which can lead to inconsistencies. On the other hand, if the
jump operator creates  a quantum state of the dot localized close
to the lead, it is not clear which energy should be used in the
occupation function for the lead electrons.

The same holds for vibrations of individual chromophores, which dominantly
couple to the local excitations. However, due to excitonic coupling the energy
eigenstates are delocalized over several molecules. Again the bath interaction
requires information on spatial and energetic (to match  the vibrational
frequencies) properties of the quantum states.

Our phenomenological PERLind approach is based on the concept of
  associating the general jump operators  $\Lj$ in Eq.~(\ref{EqLindblad}) with
  specific physical processes due to the bath coupling. Here the following
  general procedure is proposed to include both information on locality and
  the energy spectrum of the bath into the jump operators:
\begin{enumerate}
\item[\textbf{1.}] Identify the relevant transitions (numbered by $j$) in the
  system due to the bath coupling and quantify them by: (i) a dimensionless
  operator $\Lj$ specifying the change in the system and taking into account
  the spatial structure of the bath interaction;  (ii) a real dimensionless
  energy-dependent function $f_j(E)$, where $E$ is the energy the system
receives from the bath (this contains the Fermi-Dirac or Bose-Einstein
distribution for the bath excitations as well as further spectral
  properties); and (iii) a prefactor $\Gamj$, so that  $R_{\mr{i}\to
  \mr{f}}^{j}=\Gamj
f_j(E_{\mr{f}}-E_{\mr{i}})\abs{\bra{\mr{f}}\Lj\ket{\mr{i}}}^2\rmhbar{/\hbar}$
is the transition rate between the initial state $\ket{\mr{i}}$ to the final
state $\ket{\mr{f}}$.
Here, $R_{\mr{i}\to \mr{f}}^{j} $ is evaluated by Fermi's golden rule from the
microscopic bath coupling $\hat{H}_\mathrm{coup}$, where
$E_{\mr{f}}-E_{\mr{i}}$ is the energy transfer appearing in the energy balance.
\item[\textbf{2.}] Determine a basis of energy eigenstates of the system $\ket{a}$, $\ket{b}$, etc.
\item[\textbf{3.}] Represent the operators  $\Lj$ in this basis $\Ljme_{ab}=\bra{a}\Lj\ket{b}$. For particle exchange with leads, $\ket{a}$ and $\ket{b}$ have different particle numbers.
\item[\textbf{4.}] Define $\tLjme_{ab}=\Ljme_{ab}\sqrt{f_j(E_a-E_b)}$ and use the Lindblad equation
\begin{equation}\label{EqLindbladEn}
\begin{aligned}
\rmhbar{\hbar}\frac{\pd }{\pd t}\rhorme_{ab}=&\imai\bra{a}[\rhor,\hatt{H}_{\rm eff}(t)]\ket{b}
+\sum_{j,cd}\Gamj\Big(\tLjme_{ac}\rhorme_{cd}\tLjmed_{bd}\\
&-\frac{1}{2}\rhorme_{ac}\tLjmed_{dc}\tLjme_{db}
-\frac{1}{2}\tLjmed_{ca}\tLjme_{cd}\rhorme_{db}
\Big).
\end{aligned}
\end{equation}
\end{enumerate}
This procedure defines an approach for the kinetics of quantum systems in
contact with an environment. The presence of $\hatt{H}_{\mathrm{eff}}(t)$
allows for the inclusion of renormalization effects similar to the secular
approximation \cite{BreuerBook2006}. However, we do not utilize this feature
here.

In this context it is crucial to note, that the
energy information $f_j(E)$
is included
\emph{on the basis of the matrix elements} for the jump operators.
If these operators are not diagonal in the basis of
energy eigenstates, this differs essentially from conventional approaches
where $\hat{L}_j$ is defined in the form
$\sqrt{f_j(E_b-E_a)}|\Psi_b\rangle\langle
\Psi_a|$, see e.~g. Ref.~[\onlinecite{PershinPRB2008}]. As explained above, these
conventional approaches provide jumps towards energy eigenstates, which do not
reflect the spatial properties of the bath coupling.

Within the first step, the identification of jump operators can be tricky, if
the same bath couples to different transitions in the system. This can be
either understood as different jump processes for each transition or a
combined one, where all transitions are subsumed in one operator $\Lj$. For
several situations, we found that the result depends upon this choice -- an
example is given in \sectionname~\ref{SecDoubleDot}. Here we find consistent
results, if all transitions connected to identical degrees of freedom in the
bath are grouped to a single jump operator $\Lj$.

We note that our Eq.~(\ref{EqLindbladEn}) has the form of
Eq.~\eqref{EqTensor} with the tensor
\begin{equation}\label{EqNewTensor}
\begin{aligned}
K^\mathrm{PERLind}_{abcd}=&-\sum_{j}\Gamj
\Big(
\tLjme_{ac}\tLjmed_{bd}
-\frac{1}{2}\sum_{e}\tLjmed_{ed}\tLjme_{eb}\delta_{ac}\\
&-\frac{1}{2}\sum_{e}\tLjmed_{ea}\tLjme_{ec}\delta_{bd}
\Big).
\end{aligned}
\end{equation}
By construction, we find
%
%\[\begin{split}
\begin{equation}
\begin{aligned}
&\rmhbar{\frac{1}{\hbar}}K^\mathrm{PERLind}_{aacc}=-\sum_j\Big(R^{j}_{c\to a}-
\sum_{e} R^{j}_{a\to e}\delta_{ac}\Big),\\
%%%
&\rmhbar{\frac{1}{\hbar}}K^\mathrm{PERLind}_{abab}=-
\sum_{j}\Big(
\rmhbar{\frac{1}{\hbar}}\Gamj\tLjme_{aa}\tLjmed_{bb}
-\frac{1}{2}\sum_{e}[R^{j}_{b\to e}+R^{j}_{a\to e}]
\Big),
\end{aligned}
\end{equation}
%\end{split}\]
%
which are just the terms of the secular approximation for the Redfield
tensor. This shows, that our PERLind approach is an
extension of the well-established secular approximation, which is complemented
by further elements in $K^\mathrm{PERLind}_{abcd}$ with $E_b-E_a\neq
E_d-E_c$. We note that the imaginary part of $K^\mathrm{Red}_{abab}$
contributes to $\hatt{H}_{\mathrm{eff}}$ as a renormalization of the energies
$E_b-E_a \to E_b-E_a-\Imag(K^\mathrm{Red}_{abab})$, however, we do not
consider such terms in our approach.

Finally, we consider the equilibration of the system in the limit of
weak system-bath coupling. If all baths have the same temperature
(and chemical potential for particle exchange), we expect that
the density matrix relaxes to its equilibrium value
for $\hat{H}_\textrm{ext}(t)=0$. In  \appendixname~\ref{SecEquilibrium} we show that this is indeed the case for our approach
in the limit of small system-bath coupling.

\section{Application 1: Spin-polarized double-dot structure}\label{SecDoubleDot}

\begin{figure}
\begin{center}
\includegraphics[width=0.9\columnwidth]{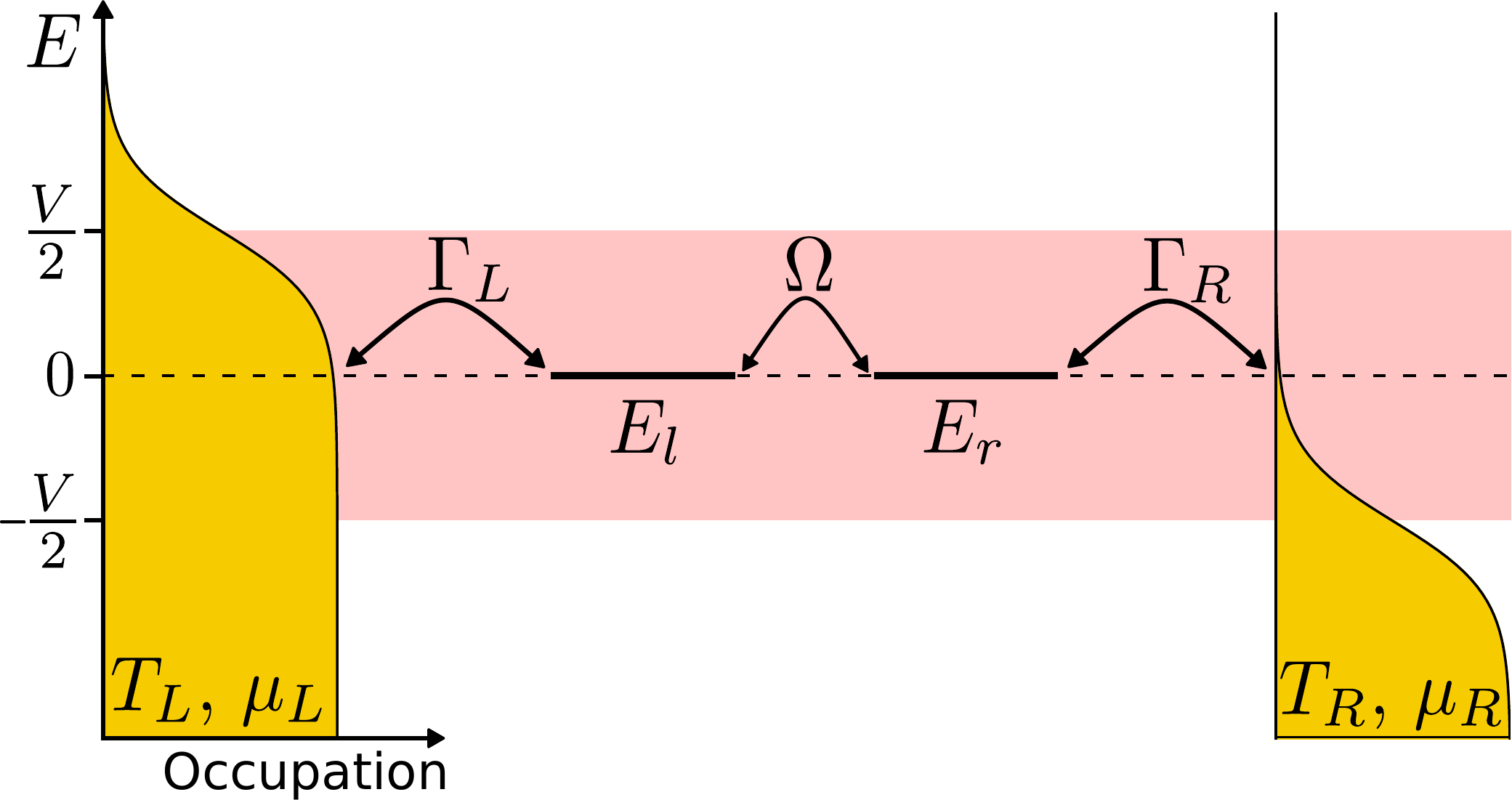}
\caption{(Color online) A simple spin-polarized double-dot structure used to practically demonstrate the  PERLind approach. The energy of the dot states is shifted by a gate voltage $V_g=E_{l}=E_{r}$. Both dots are coupled to each other ($\Omega$) and to one lead each ($\Gamma_{L}$ and $\Gamma_{R}$). The two leads are described as electron reservoirs with applied bias $V=\mu_{L}-\mu_{R}$, which results in a particle current $I$ and an energy current $\dot{E}$. }
\label{fig2}
\end{center}
\end{figure}

To demonstrate our proposed PERLind scheme, we consider a system of two
coupled quantum dots, where each dot has a single spin-polarized energy level
(indices $l$ and $r$). Both dots are coupled to each other and to source ($L$)
and drain ($R$) leads as depicted in \figurename~\ref{fig2}. We have the total
Hamiltonian\cite{GurvitzPRB1996,GurvitzPRB1998,NovotnyEPL2002,LevyEPL2014}
$\hatt{H}_S+\hatt{H}_\mr{leads}+\hatt{H}_\mr{coup}$ with the terms
\begin{subequations}
\begin{align}
\label{Hamiltonian_dot}
{\hat H}_S &= V_g\,(\dd_{l}\dan_{l} + \dd_{r}\dan_{r})
  - \Omega\, (\dd_{l}\dan_{r} + \dd_{r}\dan_{l}) \nonumber \\
&\quad{}+ U\, \dd_{l}\dan_{l}\dd_{r}\dan_{r}, \\
\label{Hamiltonian_leads}{\hat H}_\mr{leads} &= \sum\limits_{\leadqn k}
  E_{\leadqn k}^{\phantom{\dag}}\, \cd_{\leadqn k} \can_{\leadqn k}, \\
\label{Hamiltonian_hyb} {\hat H}_\mr{coup} &= \sum\limits_{k}
  (t_{L}\dd_{l}\can_{L k}+t_{R}\dd_{r}\can_{R k}) + \mr{H.c.}
\end{align}
\end{subequations}
Here, $\cd_{\leadqn k}$ creates an electron with quantum numbers $k$ in the
lead {$\leadqn\in \{L,R\}$} and $\dd_i$ creates an electron in the dot
$i\in\{l,r\}$. The coupling between left dot ($l$) and right dot ($r$) is
given by the hybridization $\Omega$ and the level energies are given by
$E_{l}$ and $E_{r}$. We assume that the level energies are in
  resonance and controlled by the
gate voltage $V_{g}=E_{l}=E_{r}$. Additionally, there can be a charging energy
$U$ when both dots are occupied. The energy dispersion in the leads is given
by $E_{\leadqn k}$ and the electrons can tunnel between dots and leads with
tunneling amplitudes $t_{L}$ and $t_{R}$. The coupling to the leads is
quantified by the transition rates $\Gamma_{\leadqn}=2\pi\sum_k
\abs{t_{\leadqn}}^2\,\delta(E-E_{\leadqn k})$, which are assumed to be
independent of the energy $E$ (so-called wide-band limit).  We also assume
that the leads are in thermal equilibrium and electron occupation is described
by a Fermi-Dirac occupation function
$f^\mathrm{FD}_\leadqn(E)=[e^{(E-\mu_{\leadqn})/T_\leadqn}+1]^{-1}$, where a bias
$V=\mu_{L}-\mu_{R}$ is applied.

Now we describe the kinetics of the reduced density matrix of the double-dot using the four steps defined in \sectionname~\ref{SecModel}:

\begin{enumerate}
\item[\textbf{1.}] There are four different tunneling processes from the leads to the dots:
\begin{enumerate}
\item[I.] An electron enters from the left lead into the left dot:
$\hatt{L}_{\rm{I}}=\dd_{l}$, $f_{\mr{I}}(E)=f^\mathrm{FD}_L(E)$, $\Gamma_{\mr{I}}=\Gamma_{L}$.
\item[II.] An electron leaves the left dot into the left lead:
$\hatt{L}_{\rm{II}}=\dan_{l}$, $f_{\mr{II}}(E)=1-f^\mathrm{FD}_L(E)$, $\Gamma_{\mr{II}}=\Gamma_{L}$.
\item[III.] An electron enters from the right lead into the right dot:
$\hatt{L}_{\rm{III}}=\dd_{r}$, $f_{\mr{III}}(E)=f^\mathrm{FD}_R(E)$, $\Gamma_{\mr{III}}=\Gamma_{R}$.
\item[IV.] An electron leaves the right dot into the right lead:
$\hatt{L}_{\rm{IV}}=\dan_{r}$, $f_{\mr{IV}}(E)=1-f^\mathrm{FD}_R(E)$, $\Gamma_{\mr{IV}}=\Gamma_{R}$.
\end{enumerate}
\item[\textbf{2.}] The system Hamiltonian $H_S$, Eq.\ \eqref{Hamiltonian_dot},  has four many-particle eigenstates,
\begin{subequations}\label{eigst}
\begin{align}
&\ket{0}, &&E_{0}=0,\\
&\ket{1}=\dd_{1}\ket{0}, &&E_{1}=V_{g}-\Omega,\\
&\ket{1'}=\dd_{1'}\ket{0}, &&E_{1'}=V_{g}+\Omega,\\
&\ket{2}=\dd_{1'}\dd_{1}\ket{0}, &&E_{2}=2V_{g}+U,
\end{align}
\end{subequations}
where
\begin{equation}\label{eigoop}
\dan_{1}=\frac{1}{\sqrt{2}}\left(\dan_{l}+\dan_{r}\right) \ \
\textrm{and} \ \
\dan_{1'}=\frac{1}{\sqrt{2}}\left(\dan_{l}-\dan_{r}\right).
\end{equation}
\item[\textbf{3.}] In the basis Eq.~\eqref{eigst} the jump operators $\hatt{L}_{j}$ are:
\begin{alignat*}{2}
L_{\mr{I}}=&\frac{1}{\sqrt{2}}
\begin{pmatrix}
0 & 0 & 0 & 0\\
+1 & 0 & 0 & 0\\
+1 & 0 & 0 & 0\\
0 & +1 & -1 & 0
\end{pmatrix},\quad & L_{\mr{II}}=&(L_{\mr{I}})^{T},\\
L_{\mr{III}}=&\frac{1}{\sqrt{2}}
\begin{pmatrix}
0 & 0 & 0 & 0\\
+1 & 0 & 0 & 0\\
-1 & 0 & 0 & 0\\
0 & -1 & -1 & 0
\end{pmatrix},\quad & L_{\mr{IV}}=&(L_{\mr{III}})^{T}.
\end{alignat*}
\begin{widetext}
\item[\textbf{4.}] The jump operators $\hatt{L}_{j}$ are weighted by
  $\sqrt{f_{j}(E)}$ to give
  $\tilde{L}^{j}_{ab}=L^{j}_{ab}\sqrt{f_{j}(E_{a}-E_{b})}$. Thus
\begin{alignat*}{2}
\tilde{L}_{\mr{I}}=&\frac{1}{\sqrt{2}}
\begin{pmatrix}
0 & 0 & 0 & 0\\
\sqrt{f^\mathrm{FD}_L(\vg-\Omega)} & 0 & 0 & 0\\
\sqrt{f^\mathrm{FD}_L(\vg+\Omega)} & 0 & 0 & 0\\
0 & \sqrt{f^\mathrm{FD}_L(\vg+\Omega+U)} & - \sqrt{f^\mathrm{FD}_L(\vg-\Omega+U)} & 0
\end{pmatrix},\\
\tilde{L}_{\mr{III}}=&\frac{1}{\sqrt{2}}
\begin{pmatrix}
0 & 0 & 0 & 0\\
\sqrt{f^\mathrm{FD}_R(\vg-\Omega)} & 0 & 0 & 0\\
-\sqrt{f^\mathrm{FD}_R(\vg+\Omega)} & 0 & 0 & 0\\
0 & -\sqrt{f^\mathrm{FD}_R(\vg+\Omega+U)} & -\sqrt{f^\mathrm{FD}_R(\vg-\Omega+U)} & 0
\end{pmatrix},\\
\end{alignat*}
\end{widetext}
\end{enumerate}
and $\tilde{L}_{\mr{II}}$ ($\tilde{L}_{\mr{IV}}$) is obtained by transposing $\tilde{L}_{\mr{I}}$ ($\tilde{L}_{\mr{III}}$) and by replacing $f_{\ell}^{FD}$ with $1-f_{\ell}^{FD}$.

After inserting $\Gamma_{j}$ and $\tilde{L}_{j}$ into Eq.~\eqref{EqLindbladEn}
we obtain a master equation for the reduced density matrix. In the long-time
limit a stationary state is reached, which we determine directly by setting
$\pd_t\rho_{ab}=0$ in Eq.~(\ref{EqLindbladEn}). For this stationary state we
can calculate various observables. In particular, we are interested in the
particle ($I_{L}$) and energy ($\dot{E}_{L}$, as relevant for thermoelectric applications\cite{SothmannNanotechnology2015})
currents flowing from the left lead into the system, which
are calculated by
\begin{equation}
I_{L}=\sum_{baa'}\left(
\Gamma_{\mr{I}}\tilde{L}^{\mr{I}}_{ba}\rhorme_{aa'}
\tilde{L}^{\mr{I}*}_{ba'}-\Gamma_{\mr{II}}\tilde{L}^{\mr{II}}_{ba}\rhorme_{aa'}
\tilde{L}^{\mr{II}*}_{ba'}\right),
\end{equation}
\begin{equation}
\dot{E}_{L}=\sum_{\substack{j=\mr{I},\mr{II}\\baa'}}
\Gamj\left(E_{b}-\frac{E_{a}+E_{a'}}{2}\right)\tLjme_{ba}\rhorme_{aa'}\tLjmed_{ba'}.
\end{equation}
See \appendixname~\ref{App:currents} for the definition and more details on the particle current and energy current observables.

We focus on the non-interacting case, $U=0$, where the transmission
formalism~\cite{CaroliJPhysCSolidStatePhys1971,ButtikerPRB1985,DattaBook1995}
provides an exact solution (see \appendixname~\ref{App:ddTrans}). The
analytic solution of the resulting master equation for the reduced density
matrix using the PERLind approach in the non-interacting case $U=0$ is
discussed in \appendixname~\ref{App:ddLA}.

\begin{figure*}
\begin{center}
\includegraphics[width=\textwidth]{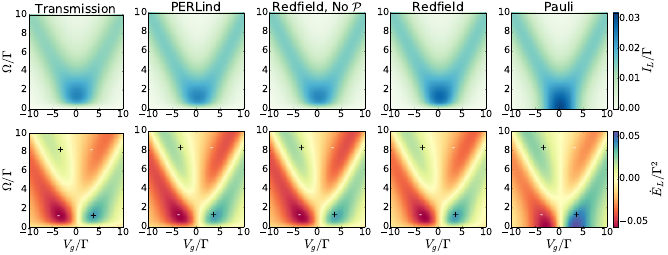}
\caption{(Color online) Simulations of the non-interacting
double-dot system using different approaches for equal energies $E_1=E_2$ and
symmetric coupling $\Gamma_{L}=\Gamma_{R}=\Gamma$. Other parameters are
$\mu_L=-\mu_R=\Gamma/4$  and $T=2\Gamma$. The +/- signs indicate the sign of the energy current. }
\label{fig3}
\end{center}
\end{figure*}

The results for symmetric coupling $\Gamma_{L}=\Gamma_{R}=\Gamma$ are shown
in Fig.~\ref{fig3}. At first we see, that the result of the PERLind
scheme is very close to the transmission result. The main difference is that
the current peaks are slightly lower and broader in the transmission
calculation, which includes tunneling to all orders and thus takes level
broadening into account. This difference vanishes with increasing
temperature, while for $T<\Gamma$ the discrepancy becomes more substantial.
We also display the result for the Redfield equations, which can be directly
applied to tunneling systems\cite{HarbolaPRB2006} (see \appendixname~\ref{App:ddRA} for more
details). This approach works reasonably well, but
agrees less with the correct transmission result than our suggested approach.
It is interesting to note that neglecting the principle value integrals
(Redfield, No $\mathcal{P}$) provides slightly better results for the used
parameters. Finally, we observe that the Pauli master equation, which has the
same stationary state as the secular approximation, fails for small interdot
coupling $\Omega\to 0$. Here the current remains finite, which is an artifact
of the neglect of coherences in the basis of eigenstates. The electrons from
the left lead enter one of the eigenstates, which are distributed over both
dots and are allowed to leave to the right lead immediately. Thus the current
is solely determined by $\Gamma$ in this case.  This issue is well-known and
appears in many circumstances~\cite{CallebautJAP2005, SchultzPRB2009, HoferNJP2017}.
It is clear that the proposed PERLind approach correctly treats this problem.
Compared to the Redfield approach, it provides results even closer to the exact
solution and, most importantly, avoids negative probabilities.
Lastly, we note that the simulations of the double dot structure
using PERLind, Redfield, and Pauli approaches were produced with the
QmeQ package \cite{KirsanskasComputPhysCommun2017} using the {\tt kerntype}
options {\tt Lindblad}, {\tt Redfield}, and  {\tt Pauli}, respectively.
This publicly available package allows to perform corresponding calculations for more complex systems in a straightforward way.

\subsection{Choice of jump operators}
The choice of jump operators, which we made so far is not a unique one. This can be seen by considering the tunneling Hamiltonian~\eqref{Hamiltonian_hyb} expressed in the eigenbasis~\eqref{eigoop}:
\begin{equation}\label{hamT_new}
\hatt{H}_\mr{coup} = \sum_{k}\left[\frac{t_{L}}{\sqrt{2}}(\dd_{1}+\dd_{1'})\can_{Lk}
+\frac{t_{R}}{\sqrt{2}}(\dd_{1}-\dd_{1'})\can_{Rk}\right]+\mr{H.c.}
\end{equation}
We can translate Eq.~\eqref{hamT_new} into jump operators at least in two different ways. Let us consider the jump processes related just to the left lead (right lead is analogous):
\begin{enumerate}
\item[\textbf{(i)}] We use four jump operators, namely $\hatt{L}_{\mr{i}}=\dd_{1}$ and  $\hatt{L}_{\mr{ii}}=\dd_{1'}$ for entering the quantum dots from the left lead as well as $\hatt{L}_{\mr{iii}}=\dan_1$ and  $\hatt{L}_{\mr{iv}}=\dan_{1'}$ for electrons leaving the quantum dots into the left lead.
\item[\textbf{(ii)}] We subsume these into two jump operators $\hatt{L}_{\mr{I}}=\dd_1+\dd_{1'}$ and $\hatt{L}_{\mr{III}}=\dan_1+\dan_{1'}$. This corresponds to the same choice which we did in the beginning of the section.
\end{enumerate}
The choice \textbf{(i)} gives the results of the Pauli master equation for the double-dot structure, where in the stationary state there are no coherences. Thus the case \textbf{(ii)} should be preferred.

A good argument for the choice \textbf{(ii)} is based on another approach. The
left lead provides electrons at a  position $z_L$ (e.~g., in the barrier
between the leftmost dot and the reservoir). Thus the jump operator is
actually the field operator $\hatt{\Psi}^\dag(z_L)$. Expanding in the state of
the quantum dot, we obtain
$\hatt{\Psi}^\dag(z_L)=\phi_1^*(z_L)\hatt{d}_1^{\dag}+\phi_{1'}^*(z_L)\hatt{d}_{1'}^{\dag}$.
Assuming an equal coupling strength for both levels in  \eqref{hamT_new} implies $\phi_1(z_L)=\phi_{1'}(z_L)$ and we obtain version \textbf{(ii)} after incorporating $\phi_1(z_L)$ into the tunneling rate $\Gamma$.

\subsection{Onsager's relation}
It was recently shown that the Redfield approach also predicts charge currents
that are not consistent with the exchange fluctuation
theorems\cite{HusseinPRB2014}, and that for our considered double-dot
structure Onsager relations relating particle current to heat current are not
satisfied (for more details see Ref.~[\onlinecite{SejaPRB2016}]). This raises
the  question: Does our proposed PERLind scheme satisfy Onsager relations? Thus we
consider the deviation $\Delta_{\mr{O}}$ from the Onsager relation for the particle current $I_{L}$ and the heat current $Q_{L}=\dot{E}_{L}-\mu_{L}I_{L}$:
\begin{equation}\label{onsrel}
\begin{aligned}
&\Delta_{\mr{O}}=L_{1}'-L_{1}=0,\\
&L_{1}'=\frac{\pd Q_{L}}{\pd V}\Big\lvert_{V=0,\Delta{T}=0}, \\
&L_{1}=T\frac{\pd I_{L}}{\pd\Delta{T}}\Big\lvert_{V=0,\Delta{T}=0}, %\quad \text{for} \quad \Delta{T}=0, \ V=0,
\end{aligned}
\end{equation}
where the bias $V$ and temperature difference $\Delta{T}$ are applied as
$\mu_{L/R}=\pm V/2$, $T_{L/R}=T\pm \Delta{T}/2$. Here, $L_{1}'$ and
$L_{1}$ are the  Onsager coefficients, which should not be confused
with the jump operators. After using analytic expressions for the currents in Eqs.~\eqref{thcur} and inserting them into relation \eqref{onsrel} we obtain for a symmetric coupling $\Gamma_{L}=\Gamma_{R}=\Gamma$
\begin{equation}\label{delta_ons}
\begin{aligned}
\Delta_{\mr{O}}=&-\frac{\Gamma^3}{32\widetilde{\Omega}T}(s+\bar{s})(f_{+}-f_{-})\\
&\times[(\bar{f}_{+}+\bar{f}_{-})s-(f_{+}+f_{-})\bar{s}]\neq 0,
\end{aligned}
\end{equation}
where
\begin{equation}\label{thnot4}
\begin{aligned}
&\widetilde{\Omega}=\Omega(1+\gamma^2),&&\gamma=\frac{\Gamma}{2\Omega},\\
&f_{\pm}=1/[\e^{(V_{g}\pm\Omega)/T_{L}}+1], &&\bar{f}_{\pm}=1-f_{\pm},\\
&s=\sqrt{f_{+}f_{-}}, &&\bar{s}=\sqrt{\bar{f}_{+}\bar{f}_{-}}.
\end{aligned}
\end{equation}
From Eq.~\eqref{delta_ons} we see that formally the Onsager's relation is not
satisfied. However, similarly as in Ref.~[\onlinecite{SejaPRB2016}] the
violation is of higher order in $\Gamma$ than the currents and for
sufficiently weak coupling no problem arises. We quantify this violation by
considering the ratio of the peak value
$\Delta_{\mr{O},\mr{peak}}=\mr{max}_{V_{g},\Omega}\abs{\Delta_{\mr{O},\mr{peak}}(V_{g},\Omega)}$
to the peak value
$L_{1,\mr{peak}}=\mr{max}_{V_{g},\Omega}\abs{L_{1}(V_{g},\Omega)}$ in the
($V_{g}$,$\Omega$) parameter space. For example, when $\Gamma=T/2$ the proposed
Lindblad scheme gives $\Delta_{\mr{O},\mr{peak}}/L_{1,\mr{peak}}\approx 0.4\%$.
The corresponding violation ratio for the Redfield approach is $16\%$ and for
the No $\mc{P}$ approach is $3\%$, which is higher than for PERLind scheme.

Alicki\cite{AlickiRepMathPhys1976} showed that the Onsager's theorem
is satisfied for a system described by Lindblad kinetics if the
\textit{quantum detailed balance} condition is fulfilled
\cite{AgarwalZPhys1973,CarmichaelZPhysBConMat1976,KossakowskiCMP1977,MajewskiJMP1984}. One
of the requirements for this is the commutation relation $[\rho,H_{S}]=0$
between the density matrix of the system and the system
Hamiltonian\cite{AlickiRepMathPhys1976, FagnolaMathNotes2008}. For our case
this implies that in the stationary state the coherences between
non-degenerate states to linear order in $V$ and $\Delta{T}$ have to be equal
to zero. This is not the case for our proposed PERLind scheme when applied to
the double-dot system, as can be seen from Eq.~\eqref{rhosols2}, so that
the \textit{quantum detailed balance} condition is violated. However, as
argued above, the non-vanishing coherence is essential to describe the
transport in the double-dot, as this provides the spatial information for
degenerate dot levels $E_r=E_l$. Thus the violation of quantum detailed balance
and Onsager's theorem is the price to pay for establishing a Lindblad type
kinetics, which provides a physically correct result.

\subsection{Asymmetric couplings}
Using asymmetric coupling $\Gamma_{L}=(1+b)\Gamma$, $\Gamma_{R}=(1-b)\Gamma$
for the two leads, we find that our proposed PERLind scheme
provides actually a current flow at zero-bias ($V=0$, $\Delta{T}=0$):
\begin{equation}
I_{L}=-\frac{1}{2}\Gamma\gamma^2b(1-b^2)(s+\bar{s})
\frac{(\bar{f}_{+}+\bar{f}_{-})s-(f_{+}+f_{-})\bar{s}}{1+\gamma^2[1-b^2(s+\bar{s})^2]}.
\end{equation}
As in the case with the violation of Onsager's theorem, this current is of
order $\Gamma^3$  and thus beyond the relevant perturbation theory.
The first-order Redfield approach also suffers from this problem,
where the current at zero-bias is determined by the principal part $\mc{P}$ integrals:
\begin{equation}
\begin{aligned}
I_{L}^{\mr{Red}}=&-\frac{\Gamma^2b(1-b^2)}{1+\gamma^2(1-b^2)}
\times\frac{1}{2\pi}\mc{P}\int_{-\infty}^{\infty}\frac{\dif{E}f(E)}{(E-V_{g})^2-\Omega^2}.
x\end{aligned}
\end{equation}
Neglecting the principal part integrals, the current becomes zero
for the Redfield approach in this particular case.

\section{Application 2: Exciton kinetics in a system of two chromophores}\label{SecPalmieri}
Here we apply the PERLind approach to exciton
  kinetics. We use the particular example discussed in
  Ref.~[\onlinecite{PalmieriJCP2009}] where a different way to  obtain a Lindblad equation from a Redfield tensor is discussed. For comparison, we show our results for the two 2-level chromophore system considered in Ref.~[\onlinecite{PalmieriJCP2009}], which is described by the total Hamiltonian $\hatt{H}_S+\hatt{H}_\mr{baths}+\hatt{H}_\mr{coup}$:
\begin{subequations}\label{tc_ham}
\begin{align}
&\label{tc_hamS}\hatt{H}_{S}=E_{1}\Bd_{1}\Ban_{1}+E_{2}\Bd_{2}\Ban_{2}+V(\Bd_{1}\Ban_{2}+\Bd_{2}\Ban_{1}),\\
&\label{tc_hamB}\hatt{H}_{\mr{baths}}=\sum_{k,i=1,2}E_{k}\ad_{ik}\aan_{ik},\\
&\label{tc_hamC}\hatt{H}_{\mr{coup}}=d_{\mr{ph}}\sum_{k,i=1,2}\Bd_{i}\Ban_{i}(\aan_{ik}+\ad_{ik}).
\end{align}
\end{subequations}
Here $\Bd_{i}$ creates an excitation on chromophore $i$, which is individually coupled to a local phonon bath. Operator $\ad_{ik}$ creates a phonon in a mode $k$ and in a bath $i$. We note that all operators satisfy canonical commutation relations $[\Ban_{i},\Bd_{i'}]=\delta_{ii'}$, $[\aan_{ik},\ad_{i'k'}]=\delta_{ii'}\delta_{kk'}$. The excitation energies are $E_1=0$ and $E_2=46.4\times hc$/cm, and the coupling strength between the excitations is $V=-71.3\times hc$/cm. The modes of the phonon baths have the density of states of over-damped Brownian oscillator~\cite{AbramaviciusChemRev2009}:
\begin{equation}
\nu(E)=\frac{2\hbar\Lambda \abs{E}}{E^2+(\hbar\Lambda)^2}\nu_{0},
\end{equation}
where $\Lambda=1/106$ fs$^{-1}$ is the inverse of the bath correlation time.

The kinetics of the reduced density matrix for the two 2-level chromophore system is described using the four steps defined in \sectionname~\ref{SecModel}:
\begin{enumerate}
\item[\textbf{1.}] There are four different jump processes, which dephases the chromophore excitations, with the same rate $\Gamma=2\pi\nu_{0}\abs{d_{\mr{ph}}}^2=2hc\lambda$, where $\lambda=35$/cm:
\begin{enumerate}
\item[I.] $\hatt{L}_{\rm{I}}=\Bd_{1}\Ban_{1}$, $f_{\mr{I}}(E)=\frac{\nu(E)}{\nu_{0}}n(E)\theta(E)$, with $n(E)=1/[\e^{E/(k_{\mr{B}}T)}-1]$, where $\theta(E)$ is the Heaviside step function.
\item[II.] $\hatt{L}_{\rm{II}}=\Bd_{1}\Ban_{1}$, $f_{\mr{II}}(E)=\frac{\nu(-E)}{\nu_{0}}[1+n(-E)]\theta(-E)$.
\item[III.] $\hatt{L}_{\rm{III}}=\Bd_{2}\Ban_{2}$, $f_{\mr{III}}(E)=\frac{\nu(E)}{\nu_{0}}n(E)\theta(E)$.
\item[IV.] $\hatt{L}_{\rm{IV}}=\Bd_{2}\Ban_{2}$, $f_{\mr{IV}}(E)=\frac{\nu(-E)}{\nu_{0}}[1+n(-E)]\theta(-E)$.
\end{enumerate}
\item[\textbf{2.}] The chromophore Hamiltonian \eqref{tc_hamS} has the excitonic eigenstates ($\hatt{H}_{S}=\ve_{1}\ed_{1}\ean_{1}+\ve_{2}\ed_{2}\ean_{2}$):
\begin{equation}\label{tc_eigoop}
\begin{aligned}
&\ean_{1}=\alpha \Ban_{1}+\beta\Ban_{2},\quad \alpha=0.81,\\
&\ean_{2}=\beta\Ban_{1}-\alpha\Ban_{2},\quad \beta=0.59,
\end{aligned}
\end{equation}
which are delocalized and which have the eigenenergies $\ve_{1}= -51.7\times hc$/cm and  $\ve_{2}=98.2\times hc$/cm.
We consider the dynamics of a single excitation in the chromophore system with the initial condition $\rho_{11}=\rho_{22}=\rho_{12}=0.5$ in the basis Eq.~\eqref{tc_eigoop}. The Hamiltonian \eqref{tc_ham} conserves the number of excitations and there is no coupling between the states with no excitation and two excitations, so it is enough to consider the subspace of a single excitation.
\item[\textbf{3.}] In the basis Eq.~\eqref{tc_eigoop} the jump operators $\hatt{L}_{j}$ are:
\begin{equation}
\begin{aligned}
&L_{\mr{I}}=
\begin{pmatrix}
\alpha^2 & \alpha\beta \\
\alpha\beta & \beta^2
\end{pmatrix},\quad L_{\mr{II}}=L_{\mr{I}},\\
&L_{\mr{III}}=
\begin{pmatrix}
\alpha^2 & -\alpha\beta \\
-\alpha\beta & \beta^2
\end{pmatrix},\quad L_{\mr{IV}}=L_{\mr{III}}.
\end{aligned}
\end{equation}
\item[\textbf{4.}] The jump operators $L_{j}$ are weighted by $\sqrt{f_{j}(E)}$ to give $\tilde{L}^{j}_{ab}=L^{j}_{ab}\sqrt{f_{j}(E_{a}-E_{b})}$.
\end{enumerate}
This provides the tensor $K^{\mr{PERLind}}$ given in Eq.~\eqref{tc_newtens}.
Its secular elements fully agree with the full Redfield tensor given in Eq.~\eqref{tc_redtens}.

\begin{figure}[h!]
\includegraphics[width=0.9\columnwidth]{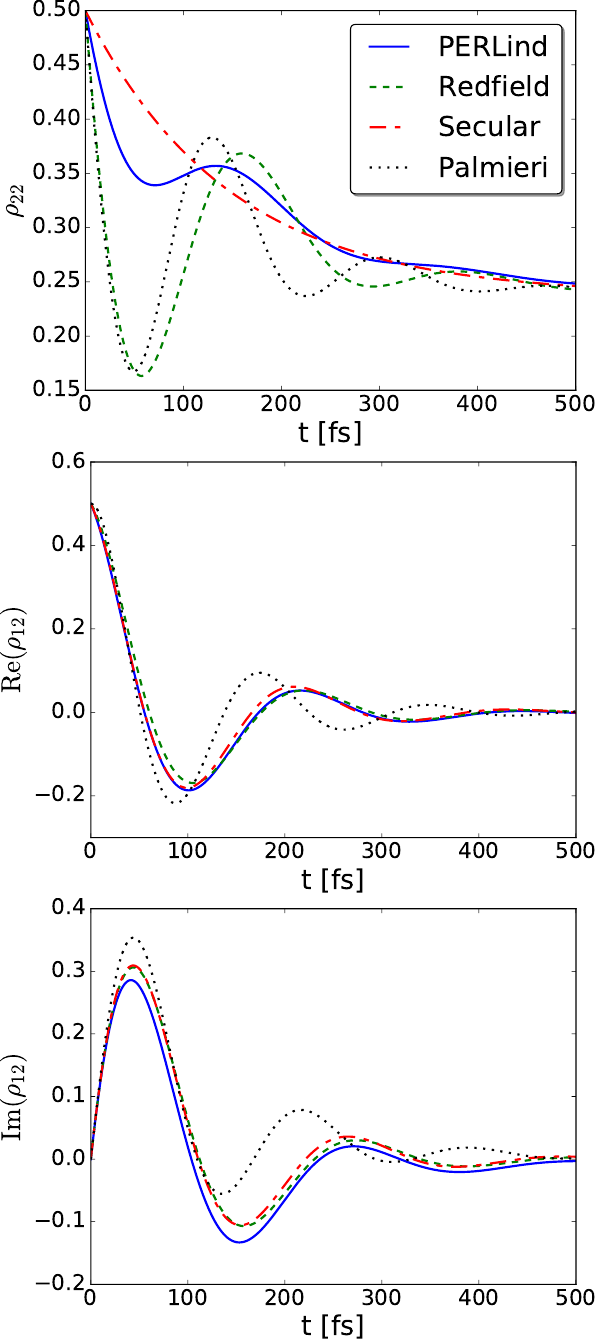}
\caption{(Color online) Results  for the system of Ref.~[\onlinecite{PalmieriJCP2009}] at $T=185$ K
and $\lambda=35$/cm. Full lines depict the results for the density matrix in
our model (Lindblad). Dashed lines show the Redfield approach, dash-dotted lines show the secular approximation, and dotted lines show the Lindblad model of Ref.~[\onlinecite{PalmieriJCP2009}] (Palmieri). The initial condition is chosen as $\rho_{11}=\rho_{22}=\rho_{12}=0.5$ in the basis of eigenstates.}
\label{fig4}
\end{figure}

In Fig.~\ref{fig4} we show the results of our PERLind approach in comparison
with the secular approximation and another Lindblad model discussed by Palmieri
\textit{et al.}\cite{PalmieriJCP2009}. We find that our approach provides
oscillating occupations, which only show a decay in the secular approximation.
However, this oscillating feature is weaker compared
to the results from the approach suggested in
Ref.~[\onlinecite{PalmieriJCP2009}] and the Redfield approach. On the other
hand, the coherences obtained from our method are closer to the Redfield
approach than the ones obtained by the previous method. As we do not have an
exact result to compare with, it is difficult to judge which method is better here.

\section{Application 3: Simulation of Quantum Cascade Lasers}\label{SecSL}

\begin{figure}
\includegraphics[width=\linewidth]{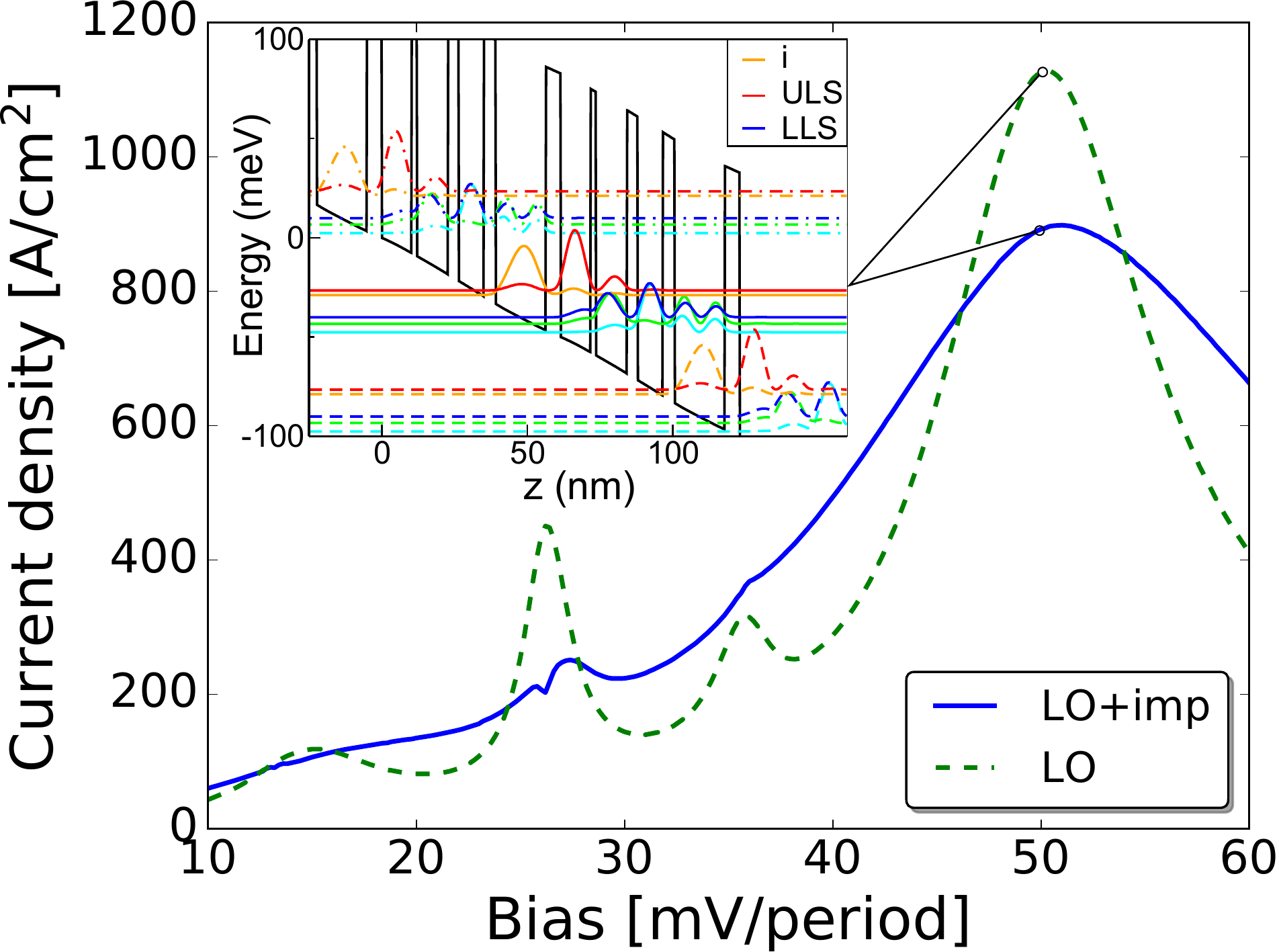}
\caption{(Color online) Current density dependence on bias per period for the QCL of Ref.~[\onlinecite{LiElectronLett2014}], with LO phonon scattering only (dashed curve), as well as with added impurity scattering (solid curve). The temperature in these simulations is $T=150$ K. The inset shows the Wannier-Stark levels at a bias of 50 mV/period (marked by circles on the current-bias curves), and the upper laser state (ULS), lower laser state (LLS), and the injector state (i) are indicated. The energy difference is 13.1 meV between the ULS and LLS, and 38.3 meV between the LLS and injector of the next period. The ULS is 1.4 meV above level i. }
\label{FigCOMETIV}
\end{figure}

The Quantum Cascade Laser (QCL)\cite{FaistBook2013,VitielloOE2015} is an
important device for the generation of infrared and terahertz
radiation. It is based on optical transitions between quantum states
$ \psi_\alpha(z)$ with energy $E_\alpha$ in the conduction band of a
semiconductor heterostructure (with growth direction $z$) as
depicted in the inset of Fig.~\ref{FigCOMETIV}. The operation relies on
an intricate interplay of tunneling and scattering transitions under the
applied bias.  Laterally, the QCL layers cover an area $A$, which is assumed
to be homogeneous (after impurity averaging) and large compared to the layer
thicknesses, so that quantization in the $x,y$ direction (expressed by bold
vectors ${\bf r}$ in the following) is not relevant.  Instead a
quasi-continuum of eigenstates $\frac{1}{\sqrt{A}}\e^{\imai {\bf k}\cdot
{\bf r}}$ with wave-vector ${\bf k}$ is assumed, so that the energy of the
quantum state $\ket{\alpha,{\bf k}}$ is $E_{\alpha \bb{k}}=E_\alpha+E_{\bb{k}}$.
Here $E_\bb{k}=\hbar^2 {\bf k}^2/(2m_c)$ with the effective mass $m_c$ of
the conduction band. Within this basis we have the single-particle
density matrix
\begin{equation}
\rho_{\beta\alpha}({\bf k})=\Tr\big\{\hat{\rho} \cd_{\alpha{\bf
        k}}\can_{\beta{\bf k}}\big\}\, .
\end{equation}
Its diagonal elements are the occupation probabilities
$f_{\alpha{\bf k}}=\rho_{\alpha\alpha}({\bf k})$. Of physical interest are
the electron densities in the individual levels (taking into account spin degeneracy)
\begin{equation}
n_\alpha=\frac{2}{A}\sum_{\bf k}f_{\alpha{\bf k}}\, ,
\end{equation}
and the current density
\begin{equation}\label{EqCurrentDens}
J(z)=\frac{-e}{A} \sum_{\bf k,\alpha\beta}
\Real\left\{\rho_{\beta\alpha}({\bf k})
\psi^*_\alpha(z)\frac{\hbar }{m_c(z)\imai}
\frac{\partial\psi_\beta(z)}{\partial z}\right\} \, ,
\end{equation}
where $e>0$ is the elementary charge.
Note that both quantities only depend on the average density matrix
\begin{equation}
\rho_{\beta\alpha}=\frac{2}{A}\sum_{\bf k}\rho_{\beta\alpha}({\bf k}).
\end{equation}
In a basis of real wavefunctions $ \psi_\alpha(z)$, such as the commonly used
energy eigenfunctions, we find that the current solely depends on the
coherences\cite{LeePRB2006}. However, these coherences
can be  approximated based on occupations\cite{CaleckiJPhysC1984} allowing
simulation schemes restricting to the occupation probabilities
$f_{\alpha{\bf k}}$
(e.g.~Refs.~[\onlinecite{TortoraPHB1999}] and [\onlinecite{JirauschekJAP2007}])
or  electron densities $n_\alpha$ (e.g.~Ref.~[\onlinecite{DonovanAPL1999}]).
Here the kinetics is given by scattering rates in the form of the
Pauli master equation. This
entirely neglects the coherences and consequently fails to describe resonant
tunneling correctly \cite{CallebautJAP2005} similar to the tunneling in the
double dot of Sec.~\ref{SecDoubleDot}.

 Considering coherences within the average density
matrix $\rho_{\alpha\beta}$ is frequently done
phenomenologically \cite{KumarPRB2009,DupontPRB2010,TerazziNJP2010}.
Taking into account the lateral degrees of freedom, more detailed
Redfield-like schemes for the simulation of $\rho_{\alpha\beta}({\bf k})$ have
been developed \cite{IottiPRL2001,WaldmullerIEEEJQuant2006,WeberPRB2009,PanPRB2017},
which can provide unphysical negative occupations as discussed
in Ref.~[\onlinecite{WeberPRB2009}].  The common solution is to use Green's function
approaches\cite{LeePRB2006,SchmielauAPL2009,KubisPRB2009,HaldasIEEEJQuant2011,WackerIEEEJSelTopQuant2013}
allowing for a full consistent treatment at the price of a high numerical
demand.
Here we show that the PERLind approach provides a viable
quantum kinetics for the average density
matrix $\rho_{\alpha\beta}$ which is based on the microscopic scattering
interaction.

An important scattering mechanism in QCLs is the electron-phonon
interaction, which enhances the electron transitions between different
subbands. For electron-phonon interaction we use the
Hamiltonian~\cite{MadelungBook1978}
\begin{equation}\label{EqVFrohlich}
\hatt{H}_{\mr{el-ph}}=\sum_{\substack{\alpha,\beta \\ \bb{k},\bb{q},q_z}}
M_{\beta\alpha}^{q_{z}}g_{\bb{q},q_z}
\cd_{\beta,\bb{k}+\bb{q}}\can_{\alpha,\bb{k}}\ban_{\bb{q},q_z}
+\mathrm{H.c.},
\end{equation}
where ${\bf q}$ and $q_z$ are the in-plane and $z$-components of the phonon
wave-vector and $\ban_{\bb{q},q_{z}}$ are the
bosonic phonon operators. Furthermore, we define
\begin{equation}
M_{\beta\alpha}^{q_z} = \int \dif{z}\,
\psi_\beta^\ast(z)e^{\imai q_zz}\psi_\alpha(z)\, .
\end{equation}
Within Fermi's golden rule this provides the scattering rates between the states in the heterostructure
\begin{equation}\label{qcl_fgr_eph}
\begin{aligned}
\Gamma&_{\alpha{\bf k}\to \beta{\bf k'}}=\frac{2\pi}{\hbar}\sum_{q_z}
|M_{\beta\alpha}^{q_z}|^2\\
&\times\Big[|g_{\bb{k}'-\bb{k},q_z}|^2\delta(E_{\beta \bb{k}'}-E_{\alpha \bb{k}}+\hbar\omega_{\mr{LO}})
(f_B(\hbar\omega_{\mr{LO}})+1)\\
&\phantom{..}
+|g_{\bb{k}-\bb{k}',-q_z}|^2\delta(E_{\beta \bb{k}'}-E_{\alpha \bb{k}}-\hbar\omega_{\mr{LO}})
f_B(\hbar\omega_{\mr{LO}})
\Big]
\end{aligned}
\end{equation}
for dispersion-less phonons with frequency $\omega_{\mr{LO}}$.
Here $f_B(E)$ is the
Bose distribution, assuming that the phonons are in thermal
equilibrium at the simulation temperature $T$.
The average transition rate for all lateral states is then given by
\begin{equation}
R_{\alpha \to \beta}=\frac{\sum_{{\bf k},{\bf k}'}f_{\alpha,{\bf k}}
\Gamma_{\alpha{\bf k}\to \beta{\bf k'}}}{\sum_{{\bf k}}f_{\alpha,{\bf k}}}.
\label{avg}
\end{equation}
Assuming that $f_{\alpha,{\bf k}}$ is a thermal distribution, this can be cast
into the form
\begin{equation}
  R_{\alpha\to \beta}=\sum_{q_z}|M_{\beta\alpha}^{q_z}|^2 f_{q_z}(E_\beta-E_\alpha).
  \label{EqStrauctureGamma}
\end{equation}
Details are given in Appendix \ref{LOScatt} for polar optical
phonon scattering. We note,
that $M_{\beta\alpha}^{q_z}$ takes into account the spatial overlap of the
states in connection with the perturbation potential, while
$f_{q_z}(E_\beta-E_\alpha)$ solely depends on the energy. This is just the form
assumed in step \textbf{1.}~of our general approach: here we identify
$M_{\beta\alpha}^{q_z}$ with $\langle\psi_\beta|\hat{L}_{q_z}|\psi_\alpha\rangle$ and
$f_{q_z}(E_\beta-E_\alpha)$ is just the distribution function, where we set
$\Gamma_{q_z}=1$. Defining
$\tilde{L}^{q_z}_{\beta\alpha}=\sqrt{f_{q_z}(E_\beta-E_\alpha)}M_{\beta\alpha}^{q_z}$
we thus obtain the  PERLind master equation (\ref{EqLindbladEn}).
The tensor (\ref{EqNewTensor}) obtained in this way
has the same structure as
Eqs.~(19) and (20) of Ref.~[\onlinecite{GordonPRB2009}], where a
slightly different notation and index labeling is used.
Going beyond these results, we also add impurity scattering in the same way
(see Appendix \ref{ImpScatt}).

We will now use this formalism to show that it can accurately simulate
QCLs. Specifically, we consider the QCL design published by Li
\textit{et al.}\cite{LiElectronLett2014} and provide a quantitative
comparison with experimental data. This design has a periodic
sequence of 180 modules, each  consisting of four
Al$_{0.16}$Ga$_{0.84}$As barriers and four GaAs wells, see the inset
of Fig.~\ref{FigCOMETIV}. As this requires far too many states to
simulate, we consider three modules together with periodic boundary
conditions, i.~e., assuming that the density matrix is identical, when
shifting all states by one period. The states $\psi_\alpha(z)$ are the
energy eigenstates which are determined following the procedure of
Ref.~[\onlinecite{LindskogSPIE2013}] and we use the five lowest states
per module (see the inset of Fig.~\ref{FigCOMETIV}), which amounts to
$15$ states in total. For our periodic conditions, the density matrix
takes into account coherences within all the states in these three central
modules as well as with
the three neighboring modules on either side.
Note that the calculations are based on the nominal experimental
sample parameters and standard semiconductor material parameters.There
is no kind of fitting.

The current density through the QCL is evaluated from
Eq.~\eqref{EqCurrentDens} by using
$\rho_{\alpha\beta}^{s}=\rho_{\alpha\beta}(+\infty)$ obtained from
stationary PERLind equations. Also we average $J(z)$ over one module
in order to compensate for spatial variations due to the finite number of
basis states. The resulting current-bias relation is shown in
Fig.~\ref{FigCOMETIV}. Taking only into account optical phonon
scattering, we find several sharp current peaks, similar to
Ref.~[\onlinecite{GordonPRB2009}]. Adding impurity scattering, these
peaks are smeared out and we observe a current peak of about 900
A/cm$^2$ in accordance with experimental
observations\cite{LiElectronLett2014}. At this operation point the
injector level is aligned with the upper laser level resulting in
efficient tunneling as shown in the inset.

\begin{figure}
\includegraphics[width=\linewidth]{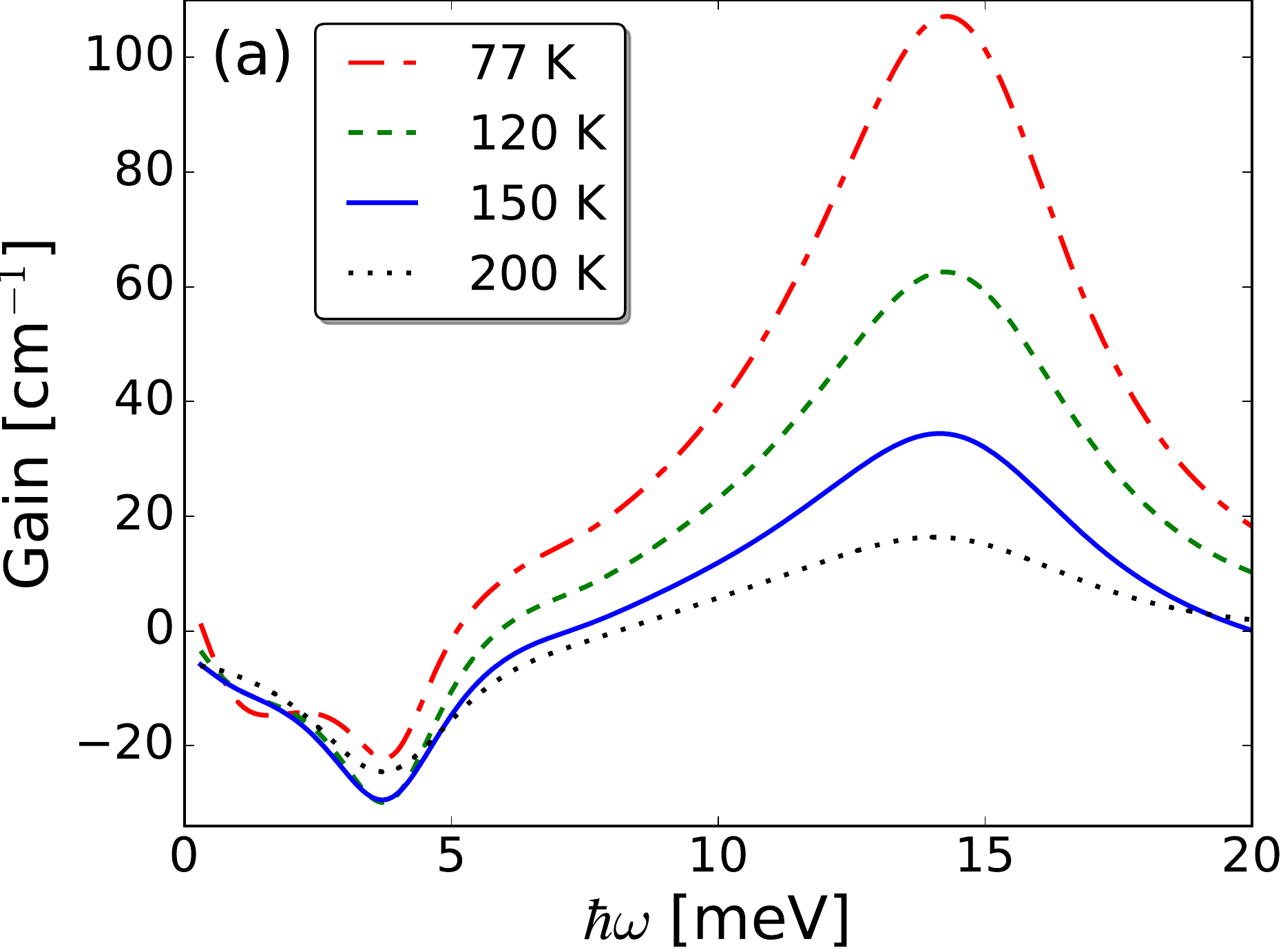}\\
\includegraphics[width=\linewidth]{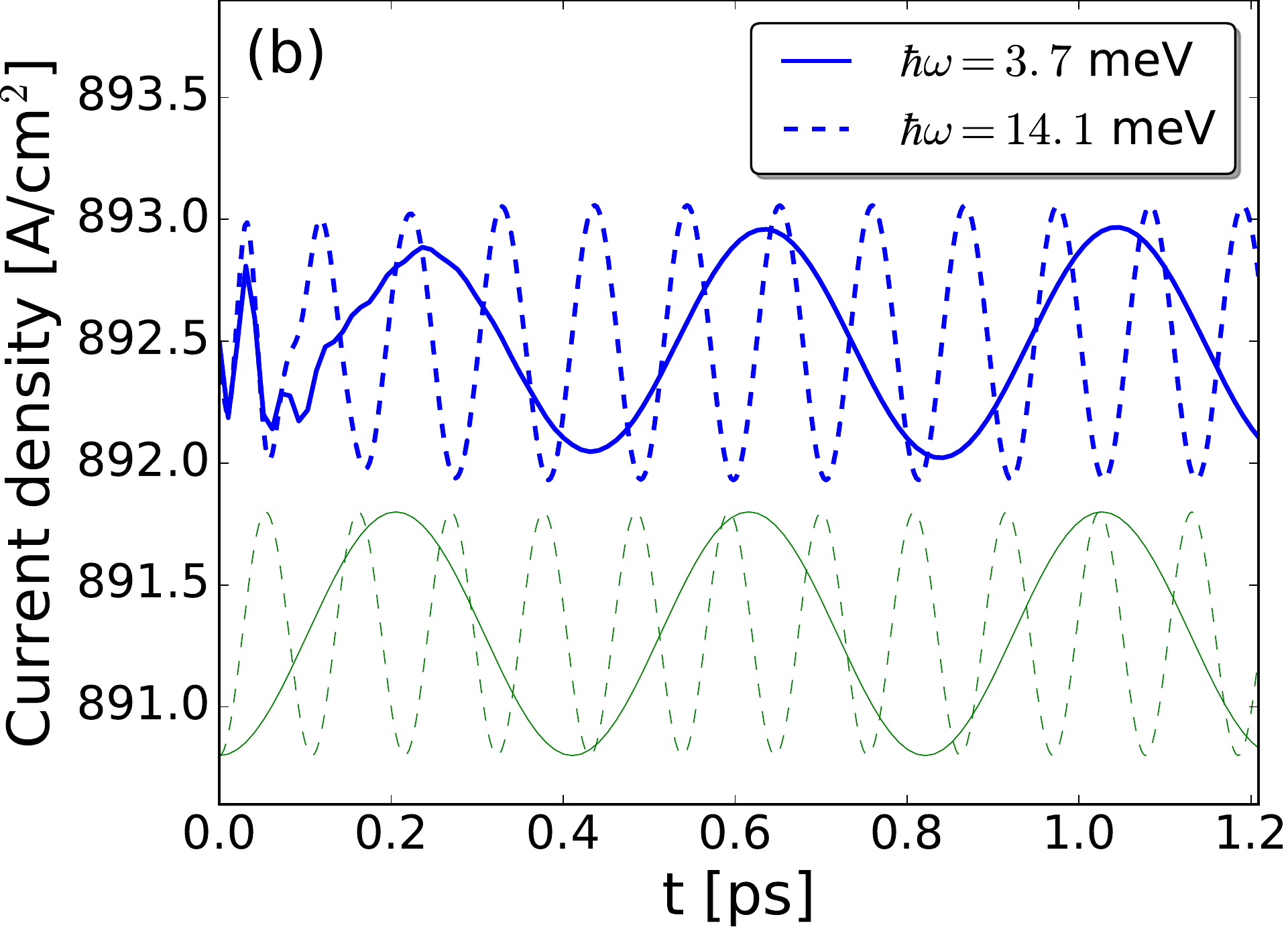}
\caption{(Color online) (a) Calculated gain spectrum for different
    simulation temperatures for the QCL design of
    Ref.~[\onlinecite{LiElectronLett2014}] at 50 mV/period bias. (b)
    Time-resolved current density at $T=150 \ \mathrm{K}$ (blue thick
    curves) and electric field (green thin curves with oscillation
    amplitude $eF_{ac}d=0.01 \ \mathrm{meV}$) at the points of maximum
    absorption ($\hbar\omega=3.7$ meV) and gain ($\hbar\omega=14.1$
    meV), where the current density is in and out of phase with the
    electric field after a transient phase, respectively.}
\label{FigCOMETgain}
\end{figure}

At a bias of 50 mV per module we observe population inversion between the
upper and lower laser states. In order to determine the optical gain,
we include the optical field (with electrical field strength $F(t)=F_\textrm{ac}\cos(\omega t)$ and frequency $\omega/2\pi$) in the model via
\begin{equation}
\hat{H}_\mathrm{ext}(t)=eF_\textrm{ac}\hat{z}\cos(\omega t).
\end{equation}
For gain simulations we evolve the PERLind equations in time, taking
the stationary density-matrix as an initial condition,
$\rho(0)=\rho^{s}$.  The field generates an alternating current
$J(t)\approx J_{\mathrm{dc}}+J_\mathrm{cos}\cos(\omega
t)+J_\mathrm{sin}\sin(\omega t)$ (after a transient phase) as depicted
in Fig.~\ref{FigCOMETgain}(b). The ratio between induced current and
ac field provides the optical gain~\cite{WackerIEEEJSelTopQuant2013}
by $-J_\mathrm{cos}/(F_\mathrm{ac}\sqrt{\epsilon_r}\epsilon_0 c)$.
This gain is positive, if  $J(t)$ and $F(t)$ are out of phase, and
conversely loss prevails  when they are in phase.
Fig.~\ref{FigCOMETgain}(a) shows the resulting gain spectra for
different simulation temperatures. We find a pronounced gain peak at
14 meV, i.e. 3.4 THz, which exactly matches the experimental
value\cite{LiElectronLett2014}.  The photon energy slightly surpasses
the separation between the upper and lower laser level (13.1 meV),
which indicates possible transitions to levels slightly below the
LLS. The gain strongly drops with temperature. For surface-plasmon
waveguides, the threshold requires gain of the order of 30-40/cm
\cite{KohenJAP2005}, which is achieved for simulation temperatures
below 150 K. Experimentally lasing was observed for heat sink
temperatures up to 123 K, which is in good accordance with our
simulations. Here, we note in passing, that the simulation temperature
should be several tens of degrees warmer than the heat sink
temperature due to non-equilibrium distributions of optical phonons
and electrons \cite{LuAPL2006b,IottiAPL2010,ShiJAP2014}, which we did
not quantify here.

Thus, the PERLind approach allows for realistic simulations
of QCLs both with respect to steady state transport and gain.
Furthermore, the  PERLind approach can also be
applied to arbitrary pulses in the optical field and
multi-mode harmonic fields containing an arbitrary number
of frequency components, which allows for a variety of interesting
applications.

\section{Conclusion}
We proposed the Position and Energy Resolving Lindblad (PERLind) approach for simulation of open quantum systems
by constructing jump operators with a specified energy and spatial dependence.
This approach combines the treatment of coherences on a microscopic basis,
such as in the Redfield kinetics, with keeping the positivity of the diagonal
elements of the density matrix. It can be
easily applied to a large variety of different physical systems, where we gave
specific examples for tunneling through quantum dot systems, exciton kinetics in
chromophores, and the simulation of quantum cascade lasers. Comparison with
the exact solution for tunneling through a double dot and experimental data
of a quantum cascade laser verifies the accuracy of the approach. On the other
hand, one has to keep in mind, that the coupling to the bath is of perturbative
nature and the projection of the system dynamics to a time-local Lindblad equation
beyond secular approximation can violate general conditions. As an example,
PERLind may violate the Onsager relations for strong bath couplings.
This appears to be the price to pay for obtaining manageable equations for
a simple description of realistic quantum systems with many degrees of freedom,
where the coherences in the steady-state are crucial.

\acknowledgments
We thank V. Trinit{\'e}, P. Hofer, and M. Hell for discussions.
Financial support from the Swedish Science Council
(Vetenskapsr{\aa}det, grant 621-2012-4024), Nano\-Lund, as well as the Knut and Alice Wallenberg
foundation are gratefully acknowledged. The figures were produced using
Matplotlib~\cite{HunterComputSciEng2007}.

\appendix

\section{Equilibrium}\label{SecEquilibrium}
Here we investigate whether thermal equilibrium $\rho^0_{ab}=\delta_{ab}\exp\left(\beta\mu N_a-\beta E_a\right)/Z$ provides a stationary solution, when all reservoirs are Bose/Fermi functions with chemical potential $\mu$ and inverse temperature $\beta=1/T$ and $\hatt{H}_{\rm eff}(t)$ is time-independent and diagonalized with the states $\ket{a},\ket{b}$, etc. We define
$\rho_{ab}=\rho^0_{ab}+\delta\rho_{ab}$ and find from Eq.~(\ref{EqLindbladEn})
\begin{equation}
\begin{aligned}
\rmhbar{\hbar}\frac{\pd }{\pd{t}}\delta\rho_{ba}
&=\imai(E_a-E_b)\delta\rho_{ba}
+\sum_{j,c}\Gamj\Big(
\tLjme_{ac}\rho^0_{cc}\tLjmed_{bc}\\
&-\frac{1}{2}\rho^0_{aa}\tLjmed_{ca}\tLjme_{cb}
-\frac{1}{2}\tLjmed_{ca}\tLjme_{cb}\rho^0_{bb}
\Big)
+{\cal O}\{\Gamma \delta\rho\}.
\end{aligned}
\end{equation}
Due to the Hermiticity of the microscopic bath couplings
(see, e.~g., Eq.~(\ref{Hamiltonian_hyb})), we find that for any matrix element
$\tLjme_{ba}$, there is a unique corresponding one with
$\tLjpme_{ab}=\tLjmed_{ba}\e^{\beta(E_b-E_a)/2-\beta\mu
  (N_b-N_a)/2}$. Here $j$ and $j'$ may result from different jump processes,
such as adding or removing a particle. We also have
$\rho_{bb}^{0}=\rho_{aa}^{0}\e^{\beta\mu(N_{b}-N_a)-\beta(E_b-E_{a})}$.
Renaming $j\to j'$ for the terms with $\tfrac{1}{2}$ we obtain
\begin{equation}
\begin{aligned}
\rmhbar{\hbar}\frac{\pd }{\pd t}\delta\rho_{ba}
&=\imai(E_a-E_b)\delta\rho_{ba}
+\sum_{j,c}\Gamj\rho^0_{cc}
\tLjme_{ac}
\tLjmed_{bc}\\
&\times\left[1-\cosh\left(\beta\frac{E_b-E_a-\mu(N_b-N_a)}{2}\right)
\right]\\
&+{\cal O}\{\Gamma \delta\rho\}.
\end{aligned}
\end{equation}
Typically the process $j$ has a defined particle exchange. Thus non-vanishing
$\tLjme_{ac}$ and $\tLjmed_{bc}$ provide $N_b=N_a$.
In the stationary state, we obtain
\begin{equation}
\begin{aligned}
\delta\rho_{ba}
&=\frac{\imai}{E_a-E_b}\sum_{j,c}\Gamj\rho^0_{cc}
\tLjme_{ac}
\tLjmed_{bc}\\
&\times\left[1-\cosh\left(\beta\frac{E_b-E_a}{2}\right)
\right]+{\cal O}\{\Gamma^2\}.
\end{aligned}
\end{equation}
Thus $\delta\rho_{ba}$ vanishes with decreasing coupling $\Gamma$. However, for $\beta|E_b-E_a|\gg 1$ the strong increase of the $\cosh$ appears to complicate the picture. As we show below, this is compensated by an exponential decay of $\rho_{cc}\tLjme_{ac}\tLjmed_{bc}$ in $\beta|E_b-E_a|$ provided the jump elements $\tilde{L}$ are bounded.

In order to show this we assume $E_b>E_a$.
We consider the state $m$ with highest occupation,
which has the effective energy $M=\textrm{ Min}_a (E_a-\mu N_a)$. Then we find
$\rho_{cc}\sim \e^{\beta(\mu N_c-E_c-M)}$. Now we assume $E_b>E_a$.
Thus $E_b-\mu N_b-M \geq E_b-E_a$. Now we consider two cases
\begin{itemize}
 \item
if $E_b-\mu N_b<E_c-\mu N_c$ then \\ $\rho_{cc}< \e^{\beta(\mu
   N_b-E_b-M)}<\e^{\beta( E_b-E_a)}$
\item
if $E_b-\mu N_b>E_c-\mu N_c$ then \\
$f_j(E_b-E_c)\sim \e^{-\beta(E_b-E_c-\mu (N_b-N_c))}$
\end{itemize}
and we find
\[\begin{split}
\rho_{cc}\tLjmed_{bc}&\lesssim \e^{\beta(\mu N_c-E_c-M)}
\e^{-\beta(E_b-E_c-\mu (N_b-N_c))/2}\Ljmed_{bc}\\
&<\e^{\beta(\mu N_c-E_c-M)/2}
\e^{-\beta(E_b-\mu N_b-M)/2}\Ljmed_{bc}\\
&<\e^{\beta(\mu N_c-E_c-M)/2}
\e^{-\beta(E_b-E_a)/2}\Ljmed_{bc}.
\end{split}\]
In both cases the exponential drop of $ \rho_{cc}\tLjme_{ac}\tLjmed_{bc}$ in $\beta(E_b-E_a)$ compensates the increase in the
$\cosh$-term. The case $E_a>E_b$ is analogous.

For the double-dot structure considered in \sectionname~\ref{SecDoubleDot} we analytically find that in the equilibrium the coherences are bounded by coupling strength $\Gamma$. From Eq.~\eqref{rhosols2} for asymmetric junction $\Gamma_{L/R}=(1\pm b)\Gamma$ we find
\begin{equation}
\rho_{11'}=\frac{b}{2}(\gamma^2+\ii\gamma)\frac{(\bar{f}_{+}+\bar{f}_{-})s-(f_{+}+f_{-})\bar{s}}{1+\gamma^2[1-b^2(s+\bar{s})^2]},
\end{equation}
which vanish for small $\Gamma$ and in this case the equilibrium is reached. Here the notation of Eq.~\eqref{thnot4} was used.

\section{\label{App:currents}Particle and energy currents}

The average particle number $\avgs{\hatt{N}}$ in the system changes by
\begin{align}\label{pdtN}
\tfrac{\pd}{\pd t}\avgs{\hatt{N}}=&\sum_{b}N_{b}\tfrac{\pd}{\pd t}\rho_{bb}\\
=&\sum_{j}\Gamj\Big(\sum_{baa'}N_{b}\tLjme_{ba}\rhorme_{aa'}\tLjmed_{ba'}\nonumber\\
&\quad-\sum_{bb'c}\frac{N_{b}}{2}[\rhorme_{bb'}\tLjmed_{cb'}\tLjme_{cb}
+\tLjmed_{cb}\tLjme_{cb'}\rhorme_{b'b}]
\Big),\nonumber
\end{align}
where $N_{b}$ denotes number of particles in the state $b$. Here we used
Eq.~\eqref{EqLindbladEn} together with the fact that the Hamiltonian
$\hatt{H}_{\mathrm{eff}}(t)$ does not change the particle
number.  Now we rename the indices $b, b', c$ by $a, a', b$ in the second term and by
$a', a, b$ in the third term of the right-hand side of Eq.~\eqref{pdtN}, which results in
\begin{equation}
\tfrac{\pd}{\pd t}\avgs{\hatt{N}}=\sum_{j,baa'}\Gamj\Big(N_{b}-\frac{N_{a}+N_{a'}}{2}\Big)
\tLjme_{ba}\rhorme_{aa'}\tLjmed_{ba'}.
\end{equation}
The jump operators $\Lj$ can be classified by the number $\Delta_{j}$ of
electrons they transfer from the leads to the system. Correspondingly, negative
$\Delta_{j}$ means the removal of particles.
Assuming that there are no coherences $\rhorme_{aa'}$ between states
with different particle number, we can replace
$N_{b}-\frac{N_{a}+N_{a'}}{2} \to \Delta_{j}$. Then all changes
due to jump operators related to transitions with lead $\leadqn$ contribute to
the current from this lead into the system:
\begin{equation}\label{elcur}
I_{\leadqn}=\sum_{\substack{j\text{ related to }\leadqn\\baa'}}
\Gamj\Delta_{j}\tLjme_{ba}\rhorme_{aa'}\tLjmed_{ba'}.
\end{equation}

Similarly we can calculate the energy current through the system. The energy of the system is defined as
\begin{equation}
E=\avgs{\hatt{H}_{S}}=\sum_{bb'}H_{bb'}\rho_{b'b}
\end{equation}
and its change is given by
\begin{align}
\tfrac{\pd}{\pd t}E=&\imai\sum_{bb'}H_{bb'}\bra{b'}[\rhor,\hatt{H}_{\mathrm{eff}}(t)]\ket{b}\nonumber\\
&+\sum_{j}\Gamj\Big(\sum_{bb'aa'}H_{bb'}\tLjme_{b'a}\rhorme_{aa'}\tLjmed_{ba'}\\
&\quad-\sum_{bb'b''c}\frac{H_{bb'}}{2}(\rhorme_{b'b''}\tLjmed_{cb''}\tLjme_{cb}
+\tLjmed_{cb'}\tLjme_{cb''}\rhorme_{b''b})
\Big).\nonumber
\end{align}
We split the first term with jump operators into two parts with exchanging $b\leftrightarrow b'$ in one of them and rename $b,b',b'',c$ by $a'',a,a',b$ in the second term and by $a'',a',a,b$ in the third term, which results in
\begin{align}\label{engrcur}
\tfrac{\pd}{\pd t}E=&P_{\mathrm{ext}}(t)
+\sum_{j,baa'}\frac{\Gamj}{2}\Big(K_{ba}^{j}\rho_{aa'}\tLjmed_{ba'}+
\tLjmed_{ba}\rho_{aa'}K_{ba'}^{j}\Big),
\end{align}
where
\begin{equation}
K_{ba}^{j}=\sum_{b'}H_{bb'}\tilde{L}_{b'a}^{j}-\sum_{a'}\tilde{L}_{ba'}^{j}H_{a'a},
\end{equation}
and
\begin{equation}
P_{\mathrm{ext}}(t)=\imai\avgs{[\hatt{H}_{S},\hatt{H}_{\mathrm{eff}}(t)]},
\end{equation}
is the power transferred to the system from the outer fields. Now if $\hatt{H}_{S}$ is diagonal (i.~e., $\hatt{H}_{S}=\sum_{b}E_{b}\ket{b}\bra{b}$) from Eq.~\eqref{engrcur} we find the energy current from the lead $\leadqn$
\begin{equation}\label{engrcur2}
\dot{E}_{\leadqn}=\sum_{\substack{j\text{ related to }\leadqn\\baa'}}
\Gamj\left(E_{b}-\frac{E_{a}+E_{a'}}{2}\right)\tLjme_{ba}\rhorme_{aa'}\tLjmed_{ba'}.
\end{equation}

\begin{widetext}
\section{Analytic solutions for the double-dot system}

\subsection{\label{App:ddTrans}Transmission formalism}
For the double-dot structure with no interactions $U=0$ the transmission formalism~\cite{CaroliJPhysCSolidStatePhys1971,ButtikerPRB1985,DattaBook1995} gives the following particle and energy currents flowing from the left lead ($L$) into the dots:
\begin{align}
\label{trans_current}
I_{L}&=\frac{1}{2\pi}\int_{-\infty}^{\infty}\dif{E}\; \mc{T}(E)\, \bigl[f_L(E) - f_R(E)\bigr],\\
\label{trans_heatcurrent}
\dot{E}_{L}&=\frac{1}{2 \pi}\int_{-\infty}^{\infty}\dif{E}\; \mc{T}(E)\, E\, \bigl[f_L(E)-f_R(E)\bigr],
\end{align}
with $f_{L/R}(E)=\left[\exp\left(\frac{E-\mu_{L/R}}{T_{L/R}}\right)+1\right]^{-1}$. For symmetric coupling $\Gamma_{L}=\Gamma_{R}=\Gamma$ the transmission function is
\begin{equation}\label{ddtrans}
\mc{T}(E)
=\biggl| \frac{\Gamma/2}{E-(V_{g}-\Omega) + \ii \Gamma /2 }- \frac{\Gamma/2}{E-(V_{g}+\Omega)+\ii \Gamma/2}\biggr|^2.
\end{equation}
We note that the above expressions are valid for the leads having an infinite bandwidth.

\subsection{\label{App:ddLA}PERLind approach}

After inserting $\Gamma_{j}$ and $\tilde{L}_{j}$ defined in \sectionname~\ref{SecDoubleDot} into Eq.~\eqref{EqLindbladEn} we obtain for non-interacting case $U=0$ such equations:
\begin{equation}\label{ddlkeq}
\pd_{t}\bm{\rho}=\mc{L}\bm{\rho},\quad
\bm{\rho}=\begin{pmatrix}
\rho_{00} & \rho_{11} & \rho_{1'1'} & \rho_{22} & \rho_{11'} & \rho_{1'1}
\end{pmatrix}^{T},
\end{equation}
\noindent with the Liouvillian $\mc{L}$
\begin{equation}
\mc{L}=
\frac{\Gamma}{2}
\begin{pmatrix}
-\Fp-\Fm & \bFm & \bFp & 0 & \bdS & \bdS \\
\Fm & -\Fp-\bFm & 0 & \bFp & \frac{1}{2}(\dS-\bdS) & \frac{1}{2}(\dS-\bdS) \\
\Fp & 0 & -\bFp-\Fm & \bFm & \frac{1}{2}(\dS-\bdS) & \frac{1}{2}(\dS-\bdS) \\
0 & \Fp & \Fm & -\bFp-\bFm & -\dS & -\dS \\
\dS & \frac{1}{2}(\dS-\bdS) & \frac{1}{2}(\dS-\bdS) & -\bdS & -2(1-\frac{\ii}{\gamma}) & 0 \\
\dS & \frac{1}{2}(\dS-\bdS) & \frac{1}{2}(\dS-\bdS) & -\bdS & 0 & -2(1+\frac{\ii}{\gamma})
\end{pmatrix}.
\end{equation}
Here we have introduced the following notations:
\begin{equation}\label{thnot}
\begin{aligned}
&\Gamma=\frac{1}{2}(\Gamma_{L}+\Gamma_{R}), \quad \gamma=\frac{\Gamma}{2\Omega},\\
&f_{L}(E)=f_{\mr{I}}(E), \quad f_{R}(E)=f_{\mr{III}}(E), \quad &&\bar{f}_{\leadqn}(E)=1-f_{\leadqn}(E),\\
&\Fpm=\frac{1}{\Gamma}[\Gamma_{L}f_{L}(\vg\pm\Omega)+\Gamma_{R}f_{R}(\vg\pm\Omega)], \quad
&&\bFpm=\frac{1}{\Gamma}[\Gamma_{L}\bar{f}_{L}(\vg\pm\Omega)+\Gamma_{R}\bar{f}_{R}(\vg\pm\Omega)],\\
&S_{\leadqn}=\frac{\Gamma_{\leadqn}}{\Gamma}\sqrt{f_{\leadqn}(\vg+\Omega)f_{\leadqn}(\vg-\Omega)}, \quad
&&\bar{S}_{\leadqn}=\frac{\Gamma_{\leadqn}}{\Gamma}\sqrt{\bar{f}_{\leadqn}(\vg+\Omega)\bar{f}_{\leadqn}(\vg-\Omega)},\\
&S_{\delta}=S_{L}-S_{R}, \quad && \bar{S}_{\delta}=\bar{S}_{L}-\bar{S}_{R}.
\end{aligned}
\end{equation}

We are interested in stationary state solution of Eq.~\eqref{ddlkeq}. By
setting $\pd_{t}\bm{\rho}=0$ and imposing normalization condition
$\Tr[\rho]=\rho_{00}+\rho_{11}+\rho_{1'1'}+\rho_{22}=1$ we obtain the solution:
\begin{equation}\label{rhosols2}
\begin{aligned}
\rho_{00}&=\frac{1}{4}\bFp\bFm-\frac{1}{8}\left[(\Fp+\Fm)\dS-(\bFp+\bFm)\bdS-4(\dS+\bdS)\right]\Real(\rho_{1'1}),\\
\rho_{11}&=\frac{1}{4}\bFp\Fm+\frac{1}{8}\left[(\Fp+\Fm)\dS-(\bFp+\bFm)\bdS\right]\Real(\rho_{1'1}),\\
\rho_{1'1'}&=\frac{1}{4}\Fp\bFm+\frac{1}{8}\left[(\Fp+\Fm)\dS-(\bFp+\bFm)\bdS\right]\Real(\rho_{1'1}),\\
\rho_{22}&=\frac{1}{4}\Fp\Fm-\frac{1}{8}\left[(\Fp+\Fm)\dS-(\bFp+\bFm)\bdS+4(\dS+\bdS)\right]\Real(\rho_{1'1}),\\
\rho_{11'}&=\frac{1}{8}\left(\gamma^2+\ii\gamma\right)\frac{(\bFp+\bFm)\dS-(\Fp+\Fm)\bdS}{1+\gamma^2\big[1-\big(\frac{\dS+\bdS}{2}\big)^2\big]},
\quad \rho_{1'1}=\rho_{11'}^{*}.
\end{aligned}
\end{equation}
Using the above expressions for the density matrix elements from Eqs.~\eqref{elcur} and \eqref{engrcur2} we get such currents
\begin{subequations}\label{thcur}
\begin{align}
&\label{cur_particle}I_{L}=\frac{1}{2}\frac{\Gamma_{L}\Gamma_{R}}{\Gamma_{L}+\Gamma_{R}}\left[g_{+}+g_{-}-(s_{L}+s_{R}+\bar{s}_L+\bar{s}_R)2\Real(\rho_{1'1})\right],\\
&\dot{E}_{L}=\frac{1}{2}\frac{\Gamma_{L}\Gamma_{R}}{\Gamma_{L}+\Gamma_{R}}\left[(\vg+\Omega)g_{+}+(\vg-\Omega)g_{-}-\vg(s_{L}+s_{R}+\bar{s}_L+\bar{s}_R)2\Real(\rho_{1'1})\right],
\end{align}
\end{subequations}
where the following notation was introduced:
\begin{equation}\label{thnot2}
\begin{aligned}
g_{\pm}=f_{L}(\vg\pm\Omega)-f_{R}(\vg\pm\Omega), \quad
s_{\leadqn}=\sqrt{f_{\leadqn}(\vg+\Omega)f_{\leadqn}(\vg-\Omega)}, \quad
\bar{s}_{\leadqn}=\sqrt{\bar{f}_{\leadqn}(\vg+\Omega)\bar{f}_{\leadqn}(\vg-\Omega)}.
\end{aligned}
\end{equation}

\subsection{\label{App:ddRA}Redfield approach}
After using Eq.~(A3) of Ref.~[\onlinecite{SejaPRB2016}] we obtain the following Liouvillian for the first-order Redfield approach:
\footnote{In non-stationary state the left hand side of Eq.~(A3) in Ref.~[\onlinecite{SejaPRB2016}] is replaced by $\ii\pd_{t}\rho_{bb'}$.}
\begin{equation}
\mc{L}_{\mr{Red}}= \frac{\Gamma}{2}
\begin{pmatrix}
-\Fp-\Fm & \bFm & \bFp & 0 & \frac{\gamma'}{\gamma}-C^{*} & \frac{\gamma'}{\gamma}-C\\
\Fm & -\Fp-\bFm & 0 & \bFp & C^{*} &   C &\\
\Fp & 0 & -\bFp-\Fm  & \bFm & C^{*} & C\\
0& \Fp & \Fm & -\bFp - \bFm & -\frac{\gamma'}{\gamma}-C^{*} & -\frac{\gamma'}{\gamma}-C\\
\frac{\gamma'}{\gamma}+C & C & C & -\frac{\gamma'}{\gamma}+C & -2(1-\frac{\ii}{\gamma})&0\\
\frac{\gamma'}{\gamma}+C^{*} & C^{*} & C^{*} & -\frac{\gamma'}{\gamma}+C^{*} &0&-2(1+\frac{\ii}{\gamma})
\end{pmatrix},
\end{equation}
where
\begin{equation}\label{thnot3}
\begin{aligned}
\gamma'&=\frac{\Gamma_{L}-\Gamma_{R}}{2\times2\Omega}, \\
C&= \frac{1}{2\pi\ii \Gamma }[(\Gamma_{L}\psi_{L+}^{*}-\Gamma_{R}\psi_{R+}^{*})-(\Gamma_{L}\psi_{L-}-\Gamma_{R}\psi_{R-})],\\
\psi_{\ell\pm}&=\Psi\left(\frac{1}{2}+\frac{\mu_{\leadqn}-(V_{g}\pm\Omega)}{\ii 2\pi T_{\leadqn}}\right).
\end{aligned}
\end{equation}
Here $\Psi(z)$ denotes the digamma function~\cite{AbramowitzBook1972}.
We also used the notations introduced in Eqs.~\eqref{thnot} and \eqref{thnot2}. For the stationary state, $\mc{L}_{\mr{Red}}\bm{\rho}=0$, we get such solution
\begin{equation}
\begin{aligned}
\rho_{00}&=\frac{1}{4}\bFp\bFm - \frac{1}{2}\Real{(C\rho_{1'1})
-\frac{1}{4}\frac{\gamma'}{\gamma}(-2-\bFm-\bFp)}\Real(\rho_{1'1}), \\
\rho_{11}&=\frac{1}{4}\Fm\bFp + \frac{1}{2}\Real{(C\rho_{1'1})}
+\frac{1}{4}\frac{\gamma'}{\gamma}(-2+\Fm+\Fp)\Real(\rho_{1'1}),\\
\rho_{1'1'}&=\frac{1}{4}\Fp\bFm + \frac{1}{2}\Real{(C\rho_{1'1})}
+\frac{1}{4}\frac{\gamma'}{\gamma}(-2+\Fm+\Fp)\Real(\rho_{1'1}),\\
\rho_{22}&=\frac{1}{4}\Fp\Fm - \frac{1}{2}\Real{(C\rho_{1'1})}
-\frac{1}{4}\frac{\gamma'}{\gamma}(+2+\Fm+\Fp)\Real(\rho_{1'1}),\\
\rho_{11'}&=\rho_{1'1}^{*}
=\frac{(\ii+\gamma)[4\gamma C-\gamma'(\Fp\Fm-\bFm\bFp)]-4\ii\gamma'^2\Imag{C}}{8(1+\gamma^2-\gamma'^2)}.
\end{aligned}
\end{equation}
The particle and energy currents are calculated using Eqs.~(A9) and (A10) of Ref.~[\onlinecite{SejaPRB2016}]:
\footnote{In Eqs.~(A9)-(A11), (B3), and (B4) of Ref.~[\onlinecite{SejaPRB2016}] the minus sign from the definition of the currents Eq.~(2) is missing. There $2\Imag$ has to be replaced by $-2\Imag$.}
\begin{align}
&I_{L}=\frac{1}{2}\frac{\Gamma_{L}\Gamma_{R}}{\Gamma_{L}+\Gamma_{R}}
\left[g_{+}+g_{-}-4\Real(\rho_{1'1})\right],\\
&\dot{E}_{L}=\frac{1}{2}\frac{\Gamma_{L}\Gamma_{R}}{\Gamma_{L}+\Gamma_{R}}
\left[(V_{g}+\Omega)g_{+}+(V_{g}-\Omega)g_{-}-4V_{g}\Real(\rho_{1'1})\right].
\end{align}
The Pauli master equation result is obtained by neglecting the coherence term $\Real(\rho_{1'1})$ in the above expressions. The result with no principal parts (Redfield, No $\mc{P}$) is obtained by neglecting the imaginary part of $C$ in Eq.~\eqref{thnot3}.

\section{Kinetic tensors for the two chromophore system}
In this \appendixname~we write down the kinetic tensors used to generate the results of \figurename~\ref{fig4}. The Redfield tensor is obtained using Eqs.~(370)-(375) of Ref.~[\onlinecite{AbramaviciusChemRev2009}]. We get slightly different numerical values than in Ref.~[\onlinecite{PalmieriJCP2009}] at $\lambda=35$/cm and $T=185$ K :
\begin{equation}\label{tc_redtens}
-K^{\mr{Red}}=
\begin{pmatrix}
-8.6 & 27.6 & 52.9 & 52.9 \\
8.6 & -27.6 & -52.9 & -52.9 \\
2.8+6.5\ii & -9.0-25.0\ii & -52.5+48.5\ii & 18.1+48.5\ii \\
2.8-6.5\ii & -9.0+25.0\ii &  18.1-48.5\ii & -52.5-48.5\ii
\end{pmatrix}\times 2\pi c/\text{cm}.
\end{equation}
Here $K^{\mr{Red}}$ corresponds to the reduced density matrix expressed as $\bm{\rho}=(\rho_{11},\rho_{22},\rho_{12},\rho_{21})^{T}$. The secular approximation is obtained by removing all the terms from $K_{abcd}^\mr{Red}$ where $E_b-E_a\neq E_d-E_c$:
\begin{equation}\label{tc_sectens}
-K^{\mr{Sec}}=
\begin{pmatrix}
-8.6 & 27.6 & 0 & 0 \\
8.6 & -27.6 & 0 & 0 \\
0 & 0 & -52.5+48.5\ii & 0 \\
0 & 0 &  0 & -52.5-48.5\ii
\end{pmatrix}\times 2\pi c/\text{cm}.
\end{equation}
We note that in simulations of \figurename~\ref{fig4} we have not used the imaginary parts of $K^{\mr{Red}}$ and $K^{\mr{Sec}}$, which corresponds to neglecting the principal part $\mc{P}$ integrals. Our proposed Lindblad scheme as discussed in \sectionname~\ref{SecPalmieri} gives:
\begin{equation}\label{tc_newtens}
-K^{\mr{PERLind}}=
\begin{pmatrix}
-8.6 & 27.6 & 17.0 & 17.0 \\
8.6 & -27.6 & -17.0 & -17.0 \\
7.4 & -26.6 & -52.5 & 15.4 \\
7.4 & -26.6 &  15.4 & -52.5
\end{pmatrix}\times 2\pi c/\text{cm}.
\end{equation}
Lastly, the procedure described in Ref.~[\onlinecite{PalmieriJCP2009}] from Eq.~\eqref{tc_redtens} gives:
\begin{equation}\label{tc_palmtens}
-K^{\mr{Palm}}=
\begin{pmatrix}
-8.6 & 27.6 & 55.1 & 55.1 \\
8.6 & -27.6 & -55.1 & -55.1 \\
-30.7 & 98.7 & -52.5 & -15.4 \\
-30.7 & 98.7 &  -15.4 & -52.5
\end{pmatrix}\times 2\pi c/\text{cm}.
\end{equation}

\end{widetext}

\section{Scattering matrix elements for QCLs}

\subsection{Electron-phonon interaction}\label{LOScatt}

 For the polar interaction with longitudinal optical phonons, the
function $g_{{\bf q},q_z}$ in the Fr{\"o}hlich Hamiltonian (\ref{EqVFrohlich}) reads~\cite{MadelungBook1978}
\begin{equation}
g_{{\bf q},q_z}=\frac{\imai}{\sqrt{AL}}\sqrt{\frac{e^2\hbar\omega_\text{LO}}
  { 2\epsilon_0\epsilon_p}}\frac{1}{\sqrt{{\bf q}^2+q_z^2}},
\end{equation}
where $L$ is the normalization length determining the $q_z$-spacing,
$\omega_\text{LO}$ is the longitudinal optical phonon frequency, which is assumed to be constant, and $\epsilon_{0}$ is the vacuum permittivity. Also $\epsilon_{p}^{-1}=\epsilon_{\infty}^{-1}-\epsilon_{s}^{-1}$, where $\epsilon_{\infty}$ and $\epsilon_{s}$ are the relative dielectric constants evaluated far above and far below $\omega_\text{LO}$, respectively.

Now we evaluate Eq.~\eqref{avg} for the emission process.
From the first term of Eq.~\eqref{qcl_fgr_eph} we find
\begin{equation}\label{emrtkkp}
\begin{aligned}
&\Gamma^\text{em.}_{\alpha\bb{k}\rightarrow\beta\bb{k}+\bb{q}}
=\frac{2\pi}{\hbar}\frac{1}{A}\frac{e^2\hbar\omega_\text{LO}}{2\epsilon_0\epsilon_p}[f_B(\hbar\omega_\text{LO})+1]\\
&\quad\times
\int\frac{\dif{q_z}}{2\pi}\frac{|M_{\beta\alpha}^{q_z}|^2}{\bb{q}^2+q_z^2}\delta(\Delta_{\beta\alpha,\bb{q}}+\frac{\hbar^2}{m_{c}}\bb{k}\cdot \bb{q} + \hbar\omega_\text{LO}),
\end{aligned}
\end{equation}
where we used the continuum limit
$\sum_{q_z}\rightarrow \frac{L}{2\pi}\int\dif q_z$ and introduced
$\Delta_{\beta\alpha,\bb{q}}=E_\beta-E_\alpha+E_\bb{q}.$
We assume that the in-plane electron states have thermal occupations, i.~e., $f_{\alpha,\bb{k}}\propto\e^{-E_{\bb{k}}/(k_{\mr{B}}T)}$. In such a case Eqs.~\eqref{avg} and \eqref{emrtkkp} give the following emission rate:
\begin{equation}\label{emrt}
\begin{aligned}
R_{\alpha\rightarrow\beta}^\text{em.}
=C_{-}\int\!\frac{\dif{q_z}}{2\pi}\int_{0}^{\infty}\!\!\frac{\dif{q}}{2\pi}
\frac{\exp\left[-\frac{(\Delta_{\beta\alpha,\bb{q}}+\hbar\omega_{\mr{LO}})^2}{4E_{\bb{q}}k_{\mr{B}}T}\right]}{\bb{q}^2+q_{z}^2}
|M_{\beta\alpha}^{q_z}|^2,
\end{aligned}
\end{equation}
where $C_{\pm}=\pm f_B(\pm\hbar\omega_\text{LO})\frac{2\pi}{\hbar}\frac{e^2\hbar\omega_\text{LO}}{2\epsilon_0\epsilon_p}\sqrt{\frac{m_{c}}{2\pi\hbar^2 k_{\mr{B}}T}}$.
Here we performed the $\bb{k}$-sums using the continuum limit $\sum_{\bb{k}}\rightarrow \frac{A}{(2\pi)^2}\int \dif^2{k}=\frac{A}{(2\pi)^2}\int_{0}^{2\pi}\dif{\phi}\int_{0}^{+\infty}\dif{k}k$ and by identifying $\bb{k}\cdot\bb{q}=kq\cos(\phi)$. Also the following integral was used:
\begin{equation}
\int_{0}^{2\pi}\dif{\phi}\sec^2(\phi)\e^{-a\sec^2(\phi)}\stackrel{a>0}{=}2\e^{-a}\sqrt{\frac{\pi}{a}}.
\end{equation}
The emission rate Eq.~\eqref{emrt} has the form of Eq.~(\ref{EqStrauctureGamma}) with
\begin{equation}
f^\text{em.}_{q_{z}}(E_\beta-E_\alpha)=
\frac{C_{-}}{L}\int_{0}^{\infty}\!\!\frac{\dif{q}}{2\pi}
\frac{\exp\left[-\frac{(\Delta_{\beta\alpha,\bb{q}}+\hbar\omega_{\mr{LO}})^2}{4E_{\bb{q}}k_{\mr{B}}T}\right]}{\bb{q}^2+q_{z}^2},
\end{equation}
which agrees with the result found by Gordon and Majer\cite{GordonPRB2009} up to a factor of 2.

For phonon absorption, we need to change the sign of $\hbar\omega_\text{LO}$ in the delta
function and change $f_B(\hbar\omega_\text{LO})+1\to f_B(\hbar\omega_\text{LO})$, resulting in
\begin{equation}
f^\text{abs.}_{q_{z}}(E_\beta-E_\alpha)=
\frac{C_{+}}{L}\int_{0}^{\infty}\!\!\frac{\dif{q}}{2\pi}
\frac{\exp\left[-\frac{(\Delta_{\beta\alpha,\bb{q}}-\hbar\omega_{\mr{LO}})^2}{4E_{\bb{q}}k_{\mr{B}}T}\right]}{\bb{q}^2+q_{z}^2}.
\end{equation}

In the simulations of Figs.~\ref{FigCOMETIV} and \ref{FigCOMETgain} we used the standard GaAs semiconductor material parameters: $m_{c}=0.067m_{e}$, $\hbar\omega_{\mr{LO}}=36.7 \ \mathrm{meV}$, $\epsilon_{s}=13.0$, $\epsilon_{\infty}=10.89$ $\epsilon_{p}=67.09$. Here $m_{e}$ denotes the mass of electron.

\subsection{Scattering by impurities}\label{ImpScatt}

For impurity scattering we have the following momentum-resolved transition rate
\begin{equation}
\Gamma_{\alpha\bb{k}\rightarrow \beta\bb{k}+\bb{q}}=\frac{2\pi}{\hbar}|\langle
U_{\beta \alpha,\bb{q}}\rangle|^2
\delta(E_{\beta,\bb{k}+\bb{q}}-E_{\alpha,\bb{k}}),
\end{equation}
where
\begin{equation}
 |\langle U_{\beta\alpha,\bb{q}}\rangle|^2 = AN_\text{2D}\sum_i w_i V^i_{\beta\alpha,\bb{q}}V^i_{\alpha\beta,-\bb{q}}
\end{equation}
is an impurity average for electron scattering and $N_\text{2D}$ is the total impurity density per period with $w_i$ being a normalized weight function distributing it on different positions $z_i$ within each period.
Here
\begin{equation}
 V_{\beta\alpha,\bb{q}}^{i}=
-\frac{e^2}{2\epsilon_0\epsilon_{s}A}\frac{\exp\left(-i\bb{q}\cdot\bb{r}_{i}\right)}{\sqrt{q^2+\lambda^2}}
M_{\beta\alpha}^{i,q},
\end{equation}
where $\lambda$ is the inverse screening length and
\begin{equation}
M_{\beta\alpha}^{i,q} =  \int \dif{z}\,
\psi^\ast_{\beta}(z)\psi_{\alpha}(z)e^{-\sqrt{q^2+\lambda^2}|z-z_i|}\, .
\end{equation}
Using Eq.~\eqref{avg}, we perform the thermal average over the in-plane momentum ($f_{\alpha,\bb{k}}\propto\e^{-E_{\bb{k}}/(k_{\mr{B}}T)}$) and obtain
\begin{equation}\label{EqGammaImp}
\begin{aligned}
R_{\alpha\rightarrow \beta}&= \sum_{i}D\int_{0}^{\infty} \frac{\dif{q}}{2\pi}\,
|M^{i,q}_{\beta\alpha}|^2
f_{i,q}(E_{\beta}-E_{\alpha})\\
&=\sum_{i,q}|M^{i,q}_{\beta\alpha}|^2f_{i,q}(E_{\beta}-E_{\alpha}),
\end{aligned}
\end{equation}
where
\begin{equation}
\begin{aligned}
f_{i,q}(E_{\beta}-E_{\alpha})=&\frac{2\pi}{D\hbar}
w_iN_\text{2D}\left(\frac{e^2}{2\epsilon_{0}\epsilon_{s}}\right)^2\sqrt{\frac{m_{c}}{2\pi\hbar^2 k_{\mr{B}}T}}\\
&\times\frac{\exp\left[-\frac{\Delta_{\beta\alpha,q}^2}{4E_{q}k_{\mr{B}}T}\right]}{q^2+\lambda^2},
\end{aligned}
\end{equation}
where $D$ is an arbitrary length scale to get the dimensions right.
As for phonon scattering, the scattering rate in
Eq.~\eqref{EqGammaImp} can thus be generalized to the Lindblad tensors
of Eq.~\eqref{EqNewTensor} with $L^{i,q}_{ab}=M^{i,q}_{ab}$ and
distribution function $f_{i,q}(E_a-E_b)$.  For the simulations we used
$N_{\mathrm{2D}}=5.16\times 10^{10} \ \mathrm{cm}^{-2}$, five impurity
layers per period with $w_{i}=0.2$ at positions $z_{i}\in\{42.0, 44.8,
47.7, 50.6, 53.4\} \ \mathrm{nm}$, and Lindhard static screening
length of $\lambda^{-1} \in \{24.3, 30.0, 33.5, 38.5\}\ \mathrm{nm}$
at $T \in \{77, 120, 150, 200\} \ \mathrm{K}$.

%\bibliography{refs_QuantTrans}

%merlin.mbs apsrev4-1.bst 2010-07-25 4.21a (PWD, AO, DPC) hacked
%Control: key (0)
%Control: author (8) initials jnrlst
%Control: editor formatted (1) identically to author
%Control: production of article title (-1) disabled
%Control: page (0) single
%Control: year (1) truncated
%Control: production of eprint (0) enabled
%

\end{document}